\documentclass{aa}
\let\linenumbers\relax

\usepackage{caption}
\usepackage{graphicx}
\usepackage{txfonts}
\usepackage{lipsum}
\usepackage{subcaption}
\usepackage{lscape}
\usepackage{placeins}
\usepackage{multirow}
\usepackage{enumitem}
\usepackage{rotating}
\usepackage{xcolor}

\newcommand{\teff}{T_{\rm eff}}
\newcommand{\logg}{\log{g}}

\newcommand{\kms}{\rm km~s^{-1}}
\newcommand{\vphi}{v_\phi}
\newcommand{\rap}{r_{\rm ap}}
\newcommand{\masyr}{{\rm mas}~{\rm yr}^{-1}}
\newcommand{\afe}{[\alpha/{\rm Fe}]}

\newcommand{\kks}{\rm kpc~km~s^{-1}}
\newcommand{\zmax}{Z_{\rm max}}

\begin{document}

\title{Deciphering the Milky Way’s star formation at cosmic noon with high proper-motion stars}
\subtitle{A precursor to the merger-driven starburst}

\author{Deokkeun An\inst{1}\thanks{\email{deokkeun@ewha.ac.kr}}
\and Young Sun Lee\inst{2}
\and Yutaka Hirai\inst{3}
\and Timothy C.\ Beers\inst{4}}

\institute{Department of Science Education, Ewha Womans University, 52 Ewhayeodae-gil, Seodaemun-gu, Seoul 03760, Republic of Korea
\and Department of Astronomy and Space Science, Chungnam National University, Daejeon 34134, Republic of Korea
\and Department of Community Service and Science, Tohoku University of Community Service and Science, 3-5-1 Iimoriyama, Sakata, Yamagata 998-8580, Japan
\and Department of Physics and Astronomy and JINA Center for the Evolution of the Elements, University of Notre Dame, Notre Dame, IN 46556, USA}

\date{Accepted July 8, 2025}

\abstract
{
Evidence suggests that the Milky Way (MW) underwent a major collision with the Gaia--Sausage/Enceladus (GSE) dwarf galaxy around cosmic noon. While GSE has since been fully disrupted, it brought in ex situ stars and dynamically heated in situ stars into the halo. In addition, the gas-rich merger may have triggered a burst of in situ star formation, potentially giving rise to a chemically distinct stellar component.
}{
We investigated the region of phase space where stars formed during the GSE merger likely reside, and retain distinct chemical and dynamical signatures.
}{
Building on our previous investigation of metallicity ([Fe/H]) and vertical angular momentum ($L_Z$) distributions, we analysed spectroscopic samples from GALAH, APOGEE, SDSS, and LAMOST, combined with Gaia kinematics. We focused on high proper-motion stars as effective tracers of the phase-space volume likely influenced by the GSE merger. To correct for selection effects, we incorporated metallicity estimates derived from SDSS and SMSS photometry.
}{
Our analysis reveals that low-$\alpha$ stars with GSE-like kinematics exhibit bimodality in [Na/Fe] and [Al/Fe] at $-1.0 \lesssim {\rm [Fe/H]} \lesssim -0.4$. One group follows the low light-element abundances of GSE stars, while another exhibits enhanced values. These low-$\alpha$, high-Na stars have eccentric orbits but are more confined to the inner MW. Eos overlaps with a high-eccentricity subset of these stars, implying that it constitutes a smaller structure nested within the broader population. After correcting for sampling biases, we estimated a population ratio of approximately 1:10 between the low-$\alpha$, high-Na stars and the GSE debris.
}{
These results suggest that the low-$\alpha$, high-Na stars formed in a compact region, likely fuelled by gas from the GSE progenitor, analogous to clumpy star-forming clouds seen in high-redshift galaxies. Such stars may trace the first sparks of more extensive merger-driven starburst activity.
}


\maketitle

\section{Introduction}

It is now widely accepted that the Milky Way (MW) experienced a major merger with the Gaia--Sausage/Enceladus (GSE) progenitor approximately $8$--$10$ billion years ago, depositing a substantial population of ex situ stars into the halo \citep{belokurov:18,helmi:18}. In addition, observations of galaxies at cosmic noon ($z \sim 2$) show that stellar discs were already in place by this time \citep{forsterschreiber:09,stott:16,lian:24,liu:24,tsukui:25,xiang:25}, with similar evidence emerging from studies of the oldest stars in the MW \citep[e.g.][]{conroy:22,xiang:25}. These findings suggest that the merger scattered a portion of in situ stars from the primordial disc into the halo under dynamically intense and turbulent conditions. This population of dynamically heated in situ stars in the MW is commonly recognised as the Splash \citep{bonaca:17,dimatteo:19,belokurov:20}.

Yet, our understanding of the full consequences of this likely gas-rich merger for the MW's evolutionary history is far from complete. Rapid gas accretion during the merger, along with tidal torques, may have triggered vigorous star formation by compressing gas to high densities and driving shocks \citep[e.g.][]{mihos:96,barnes:96}. If so, stars formed in this environment may retain distinct chemical and kinematic signatures that distinguish them from those accreted or dynamically heated. Identifying these newly formed stars and decoding their signatures in today’s stellar populations are therefore crucial steps towards reconstructing the MW’s early assembly history and clarifying the significance of the GSE merger in the context of galaxy formation and evolution.

In support of this picture, \citet{paper4} present preliminary evidence for such merger-driven star formation, based on phase-space diagrams constructed using photometric metallicity estimates and Gaia astrometry \citep{gaia:dr3}. According to the analysis, stars within the local volume ($<3$~kpc) exhibit a scale-height distribution in the [Fe/H] versus rotational velocity ($\vphi$)\footnote{$\vphi$ represents the rotational velocity in Galactocentric cylindrical coordinates in the rest frame of the MW.} plane that appears to be sculpted by a broad, valley-like feature. This distinctive pattern -- effectively traced by a coherent assembly of high proper-motion stars -- stays close to $\vphi \sim 0\ \kms$ at low metallicities ([Fe/H] $\la -1.0$), but rises sharply to $\sim180\ \kms$ over the narrow interval $-1.0 \lesssim {\rm [Fe/H]} \lesssim -0.5$. The reduced scale heights relative to the surrounding volume are a telltale sign of the GSE merger, which occurred through a low‑inclination, radial collision \citep[e.g.][]{naidu:21}. This implies that some of the stars along the valley likely originated from the same merger event, as suggested by their phase-space overlap with both the accreted GSE and dynamically heated in situ stars.

Motivated by the identification of this high proper-motion sequence (HPMS; originally termed the `Galactic starburst sequence'), \citet{paper4} propose that a portion of its metal-rich segment resulted from enhanced star formation triggered by a gas-rich merger. This interpretation is partly supported by recent numerical simulations of MW-like galaxies by \citet{grand:20}, who demonstrate that stars formed during merger-driven starbursts are relatively metal-rich ($-1.0 \lesssim {\rm [Fe/H]} \lesssim -0.5$) and span a broad range of rotational velocities ($0 \lesssim \vphi \lesssim 200\ \kms$), significantly overlapping with stars scattered from the primordial disc. Notably, this region of phase space aligns with the location of the metal-rich segment of the HPMS, suggesting that some of these stars may have formed during a burst of star formation triggered by the collision between the GSE progenitor and the young MW.

Additional independent support for merger-driven star formation comes from \citet{ciuca:24}, who analyse spectroscopic data from the Apache Point Observatory Galactic Evolution Experiment \citep[APOGEE;][]{majewski:17} in combination with asteroseismic age estimates. They identified a clump of stars associated with the so-called `great Galactic starburst' phase, characterised by rapid chemical enrichment, with [Fe/H] increasing from approximately $-0.5$ to $-0.3$ over a period of 12--13 billion years (on their asteroseismic age scale, where ages of the oldest stars exceed 18 billion years). This enrichment is accompanied by a rise in [Mg/Fe], consistent with the inflow and mixing of fresh gas during a gas-rich merger event. Together, these features provide further support for the idea that such a merger played a critical role in triggering a major starburst episode in the early MW.

However, it remains unclear how effectively and to what extent these processes enhanced shocks and/or gravitational instabilities sufficient to form a new generation of stars. A key uncertainty concerns the origin of the star-forming gas clouds. If new stars formed primarily from well-mixed primordial disc material (i.e.\ an in situ origin), their chemical and kinematic properties would likely resemble those of the thick-disc population. Conversely, if the star-forming material originated from the progenitor dwarf galaxy (i.e.\ an ex situ origin), or even from inflowing gas clouds of intergalactic origin \citep{renaud:21}, the resulting stars would be expected to exhibit distinct chemical signatures -- particularly in $\alpha$-element and light-element abundances -- relative to the pre-existing MW populations. Beyond the origin of the gas, it remains uncertain where these bursts of star formation occurred within the early MW. The gas-rich disc of the primordial MW may have experienced a global enhancement in star formation, or, alternatively, star formation may have been concentrated in localised regions where turbulence and gas compression were strongest \citep{sparre:17,yu:21,renaud:22}.

To address these outstanding questions about the GSE merger and its role in the MW's evolution, this study investigated the HPMS through detailed spectroscopic analyses and refined photometric metallicity estimates. Specifically, we hypothesised that the HPMS contains a significant contribution from stars formed in merger-driven star formation, alongside accreted GSE debris and dynamically heated disc stars. To test this hypothesis, we searched for stars of merger origin that exhibit distinct chemical enrichment signatures -- including abundances of $\alpha$-elements, sodium (Na), and aluminium (Al) -- by leveraging high-resolution spectroscopic data. Through this hypothesis-driven approach, we aimed to characterise the phase-space distribution of these stars and quantify their relative fraction.

This paper is organised as follows. In Sect.~2, we introduce the HPMS and examine its presence in spectroscopic datasets. In Sect.~3, we analyse the metallicity and angular momentum distributions of HPMS stars and identify a subset of stars likely formed through merger-driven star formation, based on distinctive Na and Al enhancements. We further investigate their spatial and dynamical properties and compare them with results from cosmological zoom-in simulations. In Sect.~4, we estimate the relative fraction of this population by decomposing the metallicity distributions in spectroscopic samples, while mitigating selection biases using photometric data. In Sect.~5, we summarise our findings and discuss their implications for the MW's evolutionary history.

\section{The high proper-motion sequence (HPMS)}

\begin{figure*}
\centering
\begin{subfigure}{8cm}
    \centering
    \includegraphics[width=\textwidth]{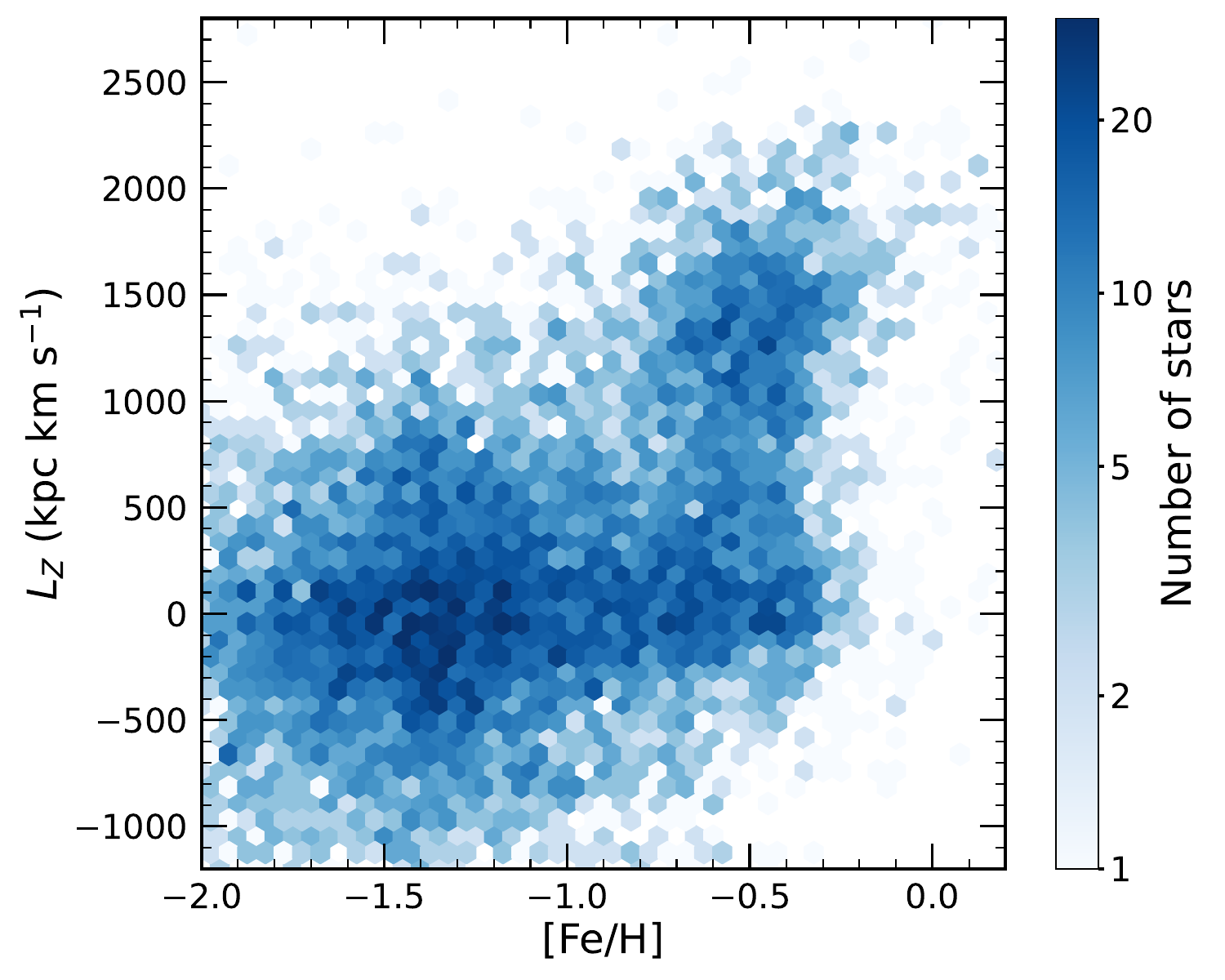}
    \caption{\textbf{Photometry}}
\end{subfigure}
\begin{subfigure}{8cm}
    \centering
    \includegraphics[width=\textwidth]{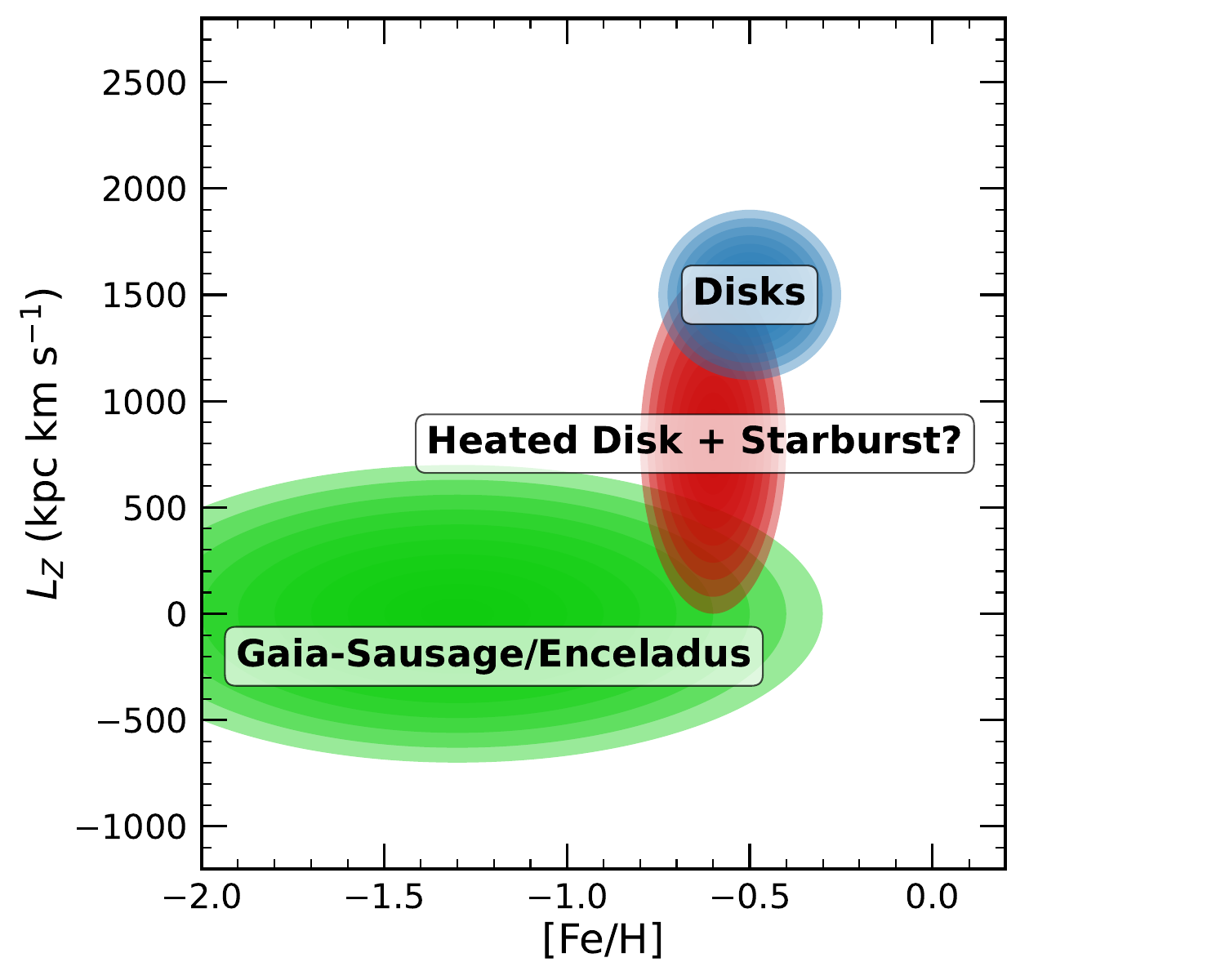}
    \caption{\textbf{Schematic view of stellar populations}}
\end{subfigure}
\begin{subfigure}{8cm}
    \centering
    \includegraphics[width=\textwidth]{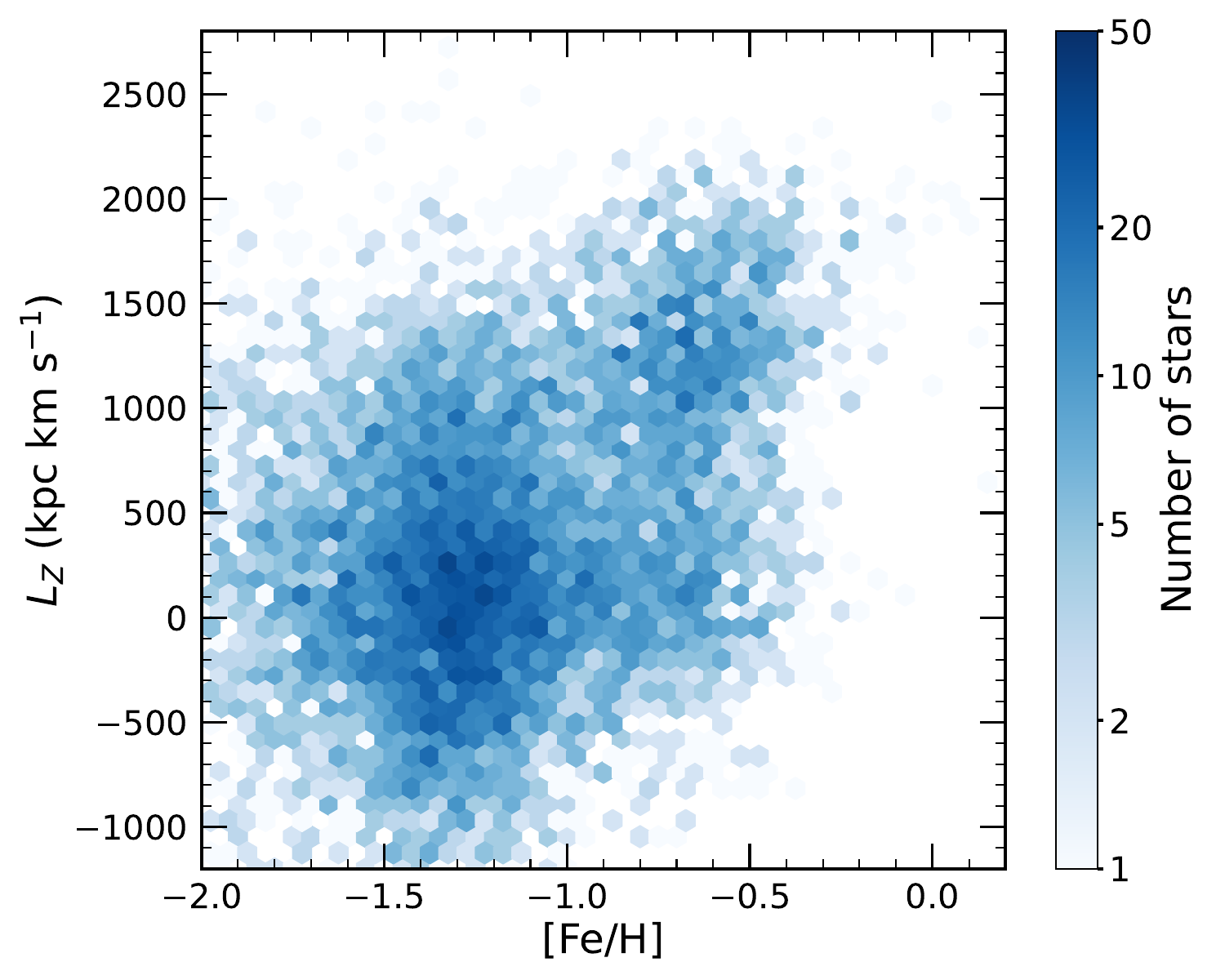}
    \caption{\textbf{SDSS}}
\end{subfigure}
\begin{subfigure}{8cm}
    \centering
    \includegraphics[width=\textwidth]{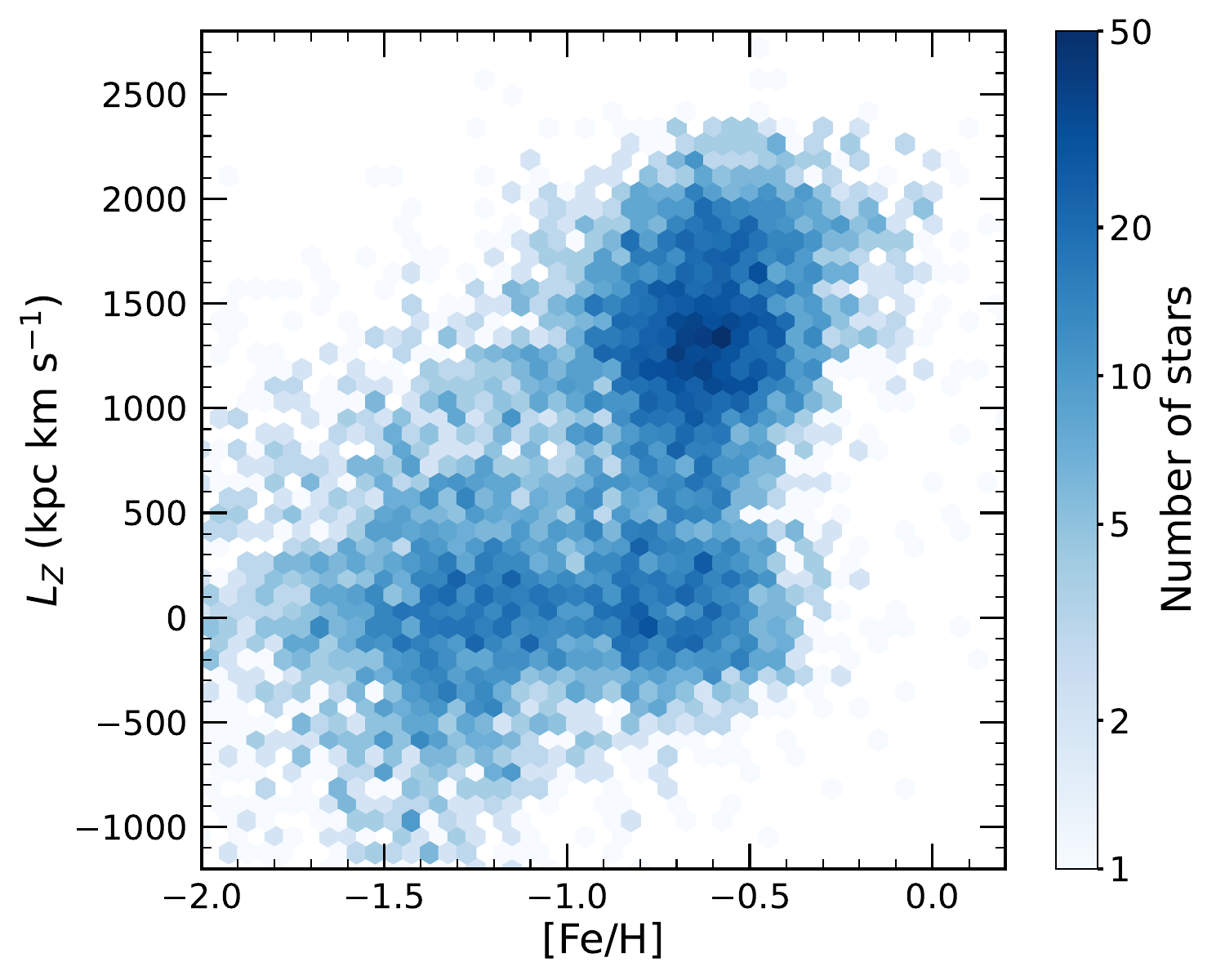}
    \caption{\textbf{LAMOST}}
\end{subfigure}
\begin{subfigure}{8cm}
    \centering
    \includegraphics[width=\textwidth]{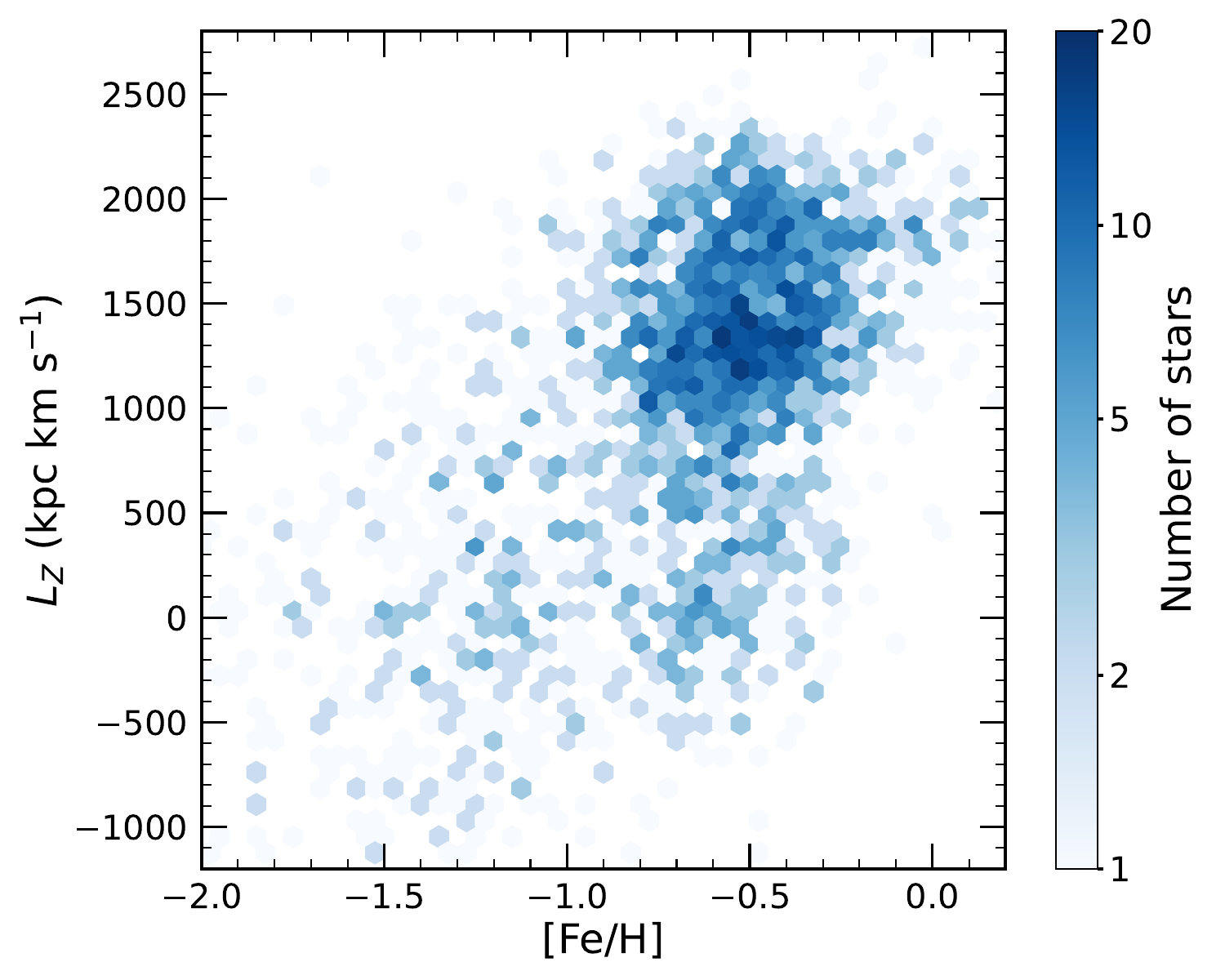}
    \caption{\textbf{GALAH}}
\end{subfigure}
\begin{subfigure}{8cm}
    \centering
    \includegraphics[width=\textwidth]{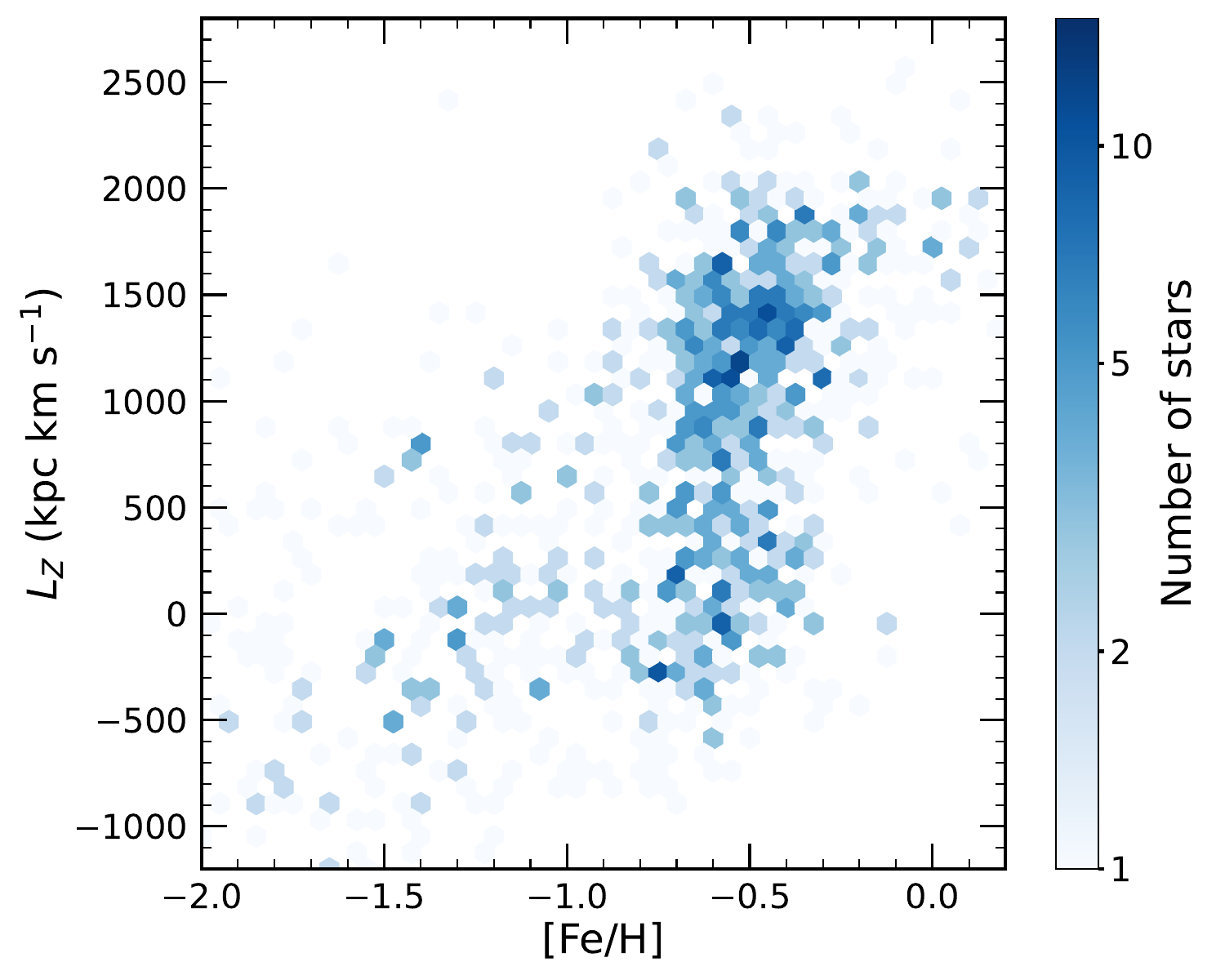}
    \caption{\textbf{APOGEE}}
\end{subfigure}
\caption{High proper-motion sequence (HPMS) in the [Fe/H]--$L_Z$ plane from various samples. Only stars with $\zmax > 2$ kpc are shown. (a) Photometric sample with proper motions exceeding $60\ \masyr$. Metallicities were estimated via Bayesian inference using SDSS and SMSS photometry, and $L_Z$ values were computed along the Galactic prime meridian without radial velocity information. (b) Schematic locations and extents of major stellar populations for reference. (c, d) Medium-resolution spectroscopic samples from SDSS and LAMOST, respectively, showing metallicities against $L_Z$ derived from full space motions. Stars with proper motions exceeding $20\ \masyr$ in SDSS and $40\ \masyr$ in LAMOST were selected. (e, f) High-resolution spectroscopic samples from GALAH and APOGEE, respectively, limited to stars with proper motions above $20\ \masyr$. For these samples, $L_Z$ values were also computed using complete kinematic information.}
\label{fig:hpms}
\end{figure*}

Figure~\ref{fig:hpms} presents the distribution of high proper-motion stars in [Fe/H] versus specific angular momentum perpendicular to the Galactic plane ($L_Z$), which forms the basis of our analysis in this study. Panel~(a) displays the distribution of stars based on photometry from the Sloan Digital Sky Survey (SDSS) Data Release (DR) 14 \citep{sdss:dr14} and the SkyMapper Southern Survey (SMSS) DR4 \citep{smss:dr4}, while the remaining panels show the [Fe/H]--$L_Z$ distributions of high proper-motion stars in large spectroscopic datasets from SDSS \citep{sdss:dr17}, the Large-sky Area Multi-Object Fiber Spectroscopic Telescope (LAMOST) DR6 \citep{cui:12,luo:15}, the GALactic Archaeology with HERMES (GALAH) DR4 \citep{buder:25}, and APOGEE DR17 \citep{majewski:17,sdss:dr17}. All datasets were cross-matched with the Gaia DR3 catalogue \citep{gaia:dr3} to obtain astrometric parameters. We included stars within 10~kpc and applied the same astrometric quality cuts across all samples, keeping only those with parallax uncertainties below 20\% and proper-motion uncertainties below 30\%. The Gaia DR3 parallaxes were corrected for the zero-point bias following the prescription of \citet{lindegren:21}. Further details on sample selection and the estimation of $L_Z$ are described below.

As is illustrated in panel~(b) of Fig.~\ref{fig:hpms}, the HPMS emerges as a coherent structure in phase space, consisting of stars with high proper motions that span a broad range in [Fe/H] and $L_Z$. Rather than representing a single stellar population, the HPMS forms a chain of previously identified components such as GSE debris \citep{belokurov:18,helmi:18,naidu:21}, the Splash \citep{bonaca:17,dimatteo:19,belokurov:20}, and disc stars, which together form a continuous sequence in this space. In the metal-poor regime, GSE stars naturally stand out in our high proper-motion samples, as their highly radial orbits result in large proper motions when observed locally, in contrast to the more isotropic motions of in situ halo stars. In the metal-rich regime, heated disc stars are also readily apparent, reflecting their more eccentric orbits compared to the nearly circular motions of disc stars.

\subsection{Photometric sample}

Panel~(a) of Fig.~\ref{fig:hpms} presents the chemo-dynamical distribution of high proper-motion ($>60\ \masyr$) stars in the [Fe/H]-$L_Z$ plane, based on SDSS DR14 and SMSS DR4 photometry. As is detailed in \citet{paper1,paper2,paper3} and \citet{paper4}, photometric metallicity estimates were obtained by matching individual stars to an empirically calibrated set of stellar isochrones for main-sequence stars ($5.0 < M_G < 7.5$), which relate photometric colours to stellar effective temperature, luminosity, and metallicity. As is shown in \citet{paper4}, metallicity maps derived from SDSS and SMSS photometry offer all-sky coverage in both the northern and southern hemispheres, excluding regions along the Galactic plane, and cover a local volume extending to approximately 3~kpc.

In this work, we adopted revised photometric metallicity estimates derived from a Bayesian approach. These estimates incorporate Gaia DR3 parallaxes and a three-dimensional extinction map \citep{paper5}\footnote{https://github.com/deokkeunan/Galactic-extinction-map} as priors. The methodology and resulting data products will be detailed in a forthcoming publication (An et al., in prep.). Compared to our previous estimates \citep{paper4}, this approach provides a more accurate treatment of unresolved binary populations within photometric catalogues, while preserving the same metallicity scale as the previous approach and maintaining consistency with the spectroscopic scale. We selected reliable [Fe/H] estimates with uncertainties of less than $0.4$~dex, focusing on stars that are likely single or are unresolved binaries with low mass ratios ($< 0.5$).

\subsection{Low-resolution spectroscopic samples}

The SDSS data were compiled from the original SDSS \citep{york:00}, the Sloan Extension for Galactic Understanding and Exploration (SEGUE; \citealt{yanny:09}; \citealt{rockosi:22}), the Baryon Oscillation Spectroscopic Survey (BOSS; \citealt{dawson:13}), and the extended Baryon Oscillation Spectroscopic Survey (eBOSS; \citealt{blanton:17}). Low-resolution stellar spectra ($R \sim 1800$) from these surveys were analysed using an updated version of the SEGUE Stellar Parameter Pipeline (SSPP; \citealt{allende:08,lee:08a,lee:08b,lee:11,smolinski:11,lee:13}). The SSPP was used to determine stellar atmospheric parameters, including effective temperature ($\teff$), surface gravity ($\logg$), metallicity ([Fe/H]), and abundance ratios such as [$\alpha$/Fe] and [C/Fe], with typical 1$\sigma$ uncertainties of 180~K, 0.24~dex, 0.23~dex, less than 0.1~dex, and less than 0.3~dex, respectively.

As is demonstrated by \citet{lee:15}, the SSPP has been successfully applied to LAMOST DR6\footnote{https://www.lamost.org/lmusers/}, owing to its similar wavelength coverage (3800--9000~\AA) and spectral resolution ($R \sim 1800$) to those of SDSS. In this study, we used stellar atmospheric parameters and elemental abundance ratios -- including [Fe/H] and [$\alpha$/Fe] -- derived with the SSPP from approximately 6 million spectra in LAMOST DR6. Among $\sim$44,000 stars common to both SDSS and LAMOST, the systematic differences are small: $5$~K in $\teff$, $0.04$~dex in $\logg$, $0.1$~dex in [Fe/H], and less than $0.02$~dex in [$\alpha$/Fe]. As these offsets are well within the typical measurement uncertainties, we did not apply any corrections in our analysis. We refer to the approach outlined by \citet{lee:23}, who present the combined SDSS and LAMOST dataset, although their sample is restricted to stars on highly eccentric orbits.

Panels~(c) and (d) of Fig.~\ref{fig:hpms} present high proper-motion stars from SDSS and LAMOST, respectively. These stars, selected within $4400 < \teff < 7000$~K, have well-measured [Fe/H] values, with metallicity and [$\alpha$/Fe] uncertainties of less than $0.3$ and $0.1$~dex, respectively, and a minimum signal-to-noise ratio (S/N) of $10$. For the LAMOST sample, we adopted a proper-motion threshold of $> 40\ \masyr$, which is less stringent than the $> 60\ \masyr$ cut used for the photometric sample, as the latter would lead to a substantial loss of stars. As is noted in \citet{paper4}, lowering the proper-motion threshold increases contamination from disc stars, which in turn diminishes the contrast and coherence of the HPMS structure. This trade-off is further shaped by survey-specific selection effects. Accordingly, the adopted proper-motion cut represents a compromise between maximising sample size and preserving the distinctiveness of the HPMS in phase space. For the SDSS sample, an even more inclusive threshold of $> 20\ \masyr$ was applied, owing to the smaller number of stars tracing the vertical segment of the HPMS.

\subsection{High-resolution spectroscopic samples}

The GALAH sample in panel~(e) of Fig.~\ref{fig:hpms} includes [Fe/H] estimates derived from high-resolution spectra ($R \sim 28,000$) from DR4 \citep{buder:25}. The $L_Z$ of individual stars was computed using full astrometric and radial velocity information from GAIA and GALAH. The $\alpha$-element abundances ([$\alpha$/Fe]) were computed as a weighted mean of [Mg/Fe], [Si/Fe], [Ca/Fe], and [Ti/Fe], including only measurements with quality flags not set ({\tt flag\_X\_fe}). We selected stars within the temperature range $4800 < \teff < 7000$~K that have reliable [Fe/H] and [$\alpha$/Fe] measurements (uncertainties below 0.3~dex and 0.1~dex, respectively, and no flag set in {\tt flag\_fe\_h}). Additionally, we required an S/N ({\tt snr\_px\_ccd3}) greater than 10 and ensure that measurements are not flagged in the GALAH pipeline ({\tt flag\_sp}, {\tt flag\_sp\_fit}, and {\tt flag\_red}).

As is shown in panel~(f), we also utilised accurate stellar parameter estimates, including [Fe/H] and [$\alpha$/M]\footnote{For the APOGEE sample, we used the [$\alpha$/M] values provided in the catalogue, which represent the abundance of $\alpha$-elements relative to the overall metal content.}, from APOGEE DR17 \citep{majewski:17,sdss:dr17}, selecting stars with spectra ($R \sim 22,500$) having S/N $> 10$, and requiring that the pipeline processing flag {\tt ASPCAPFLAG} is not set. To ensure reliable parameter estimates, we further restricted the sample to stars with $4400 < \teff < 7000$~K, and with uncertainties of less than $0.3$~dex in [Fe/H] and less than $0.1$~dex in [$\alpha$/M]. Given the relatively small number of high proper-motion stars along the HPMS in these high-resolution samples, we applied the same proper-motion threshold as in the SDSS sample ($> 20\ \masyr$).

\subsection{Orbital integration}

The $L_Z$ of individual stars was computed using parallaxes and rotational velocities ($v_\phi$) in the Galactocentric cylindrical coordinate system. We adopted a Galactic centre distance of $8.34$~kpc \citep{reid:14} and the Sun's vertical displacement of $20.8$~pc \citep{bennett:19}. The Sun's motion was taken as ($U$, $V$, $W$) = ($11.1$, $12.24$, $7.25$)$ ~\kms$ relative to the Local Standard of Rest, which has a circular velocity of $\vphi = 238\ \kms$ \citep{schonrich:10, schonrich:12}.

Additional orbital parameters of individual stars -- including the maximum vertical excursion from the Galactic plane ($\zmax$), orbital eccentricity ($e$) and apogalacticon ($\rap$) -- were computed using the {\tt galpy} package \citep{bovy:15}\footnote{http://github.com/jobovy/galpy}, adopting the MW gravitational potential from \citet{mcmillan:17}. In the case of the photometric sample, orbital parameters were derived using Gaia's proper motions along the Galactic prime meridian (within $\pm30^\circ$ of $l = 0^\circ$ and $l = 180^\circ$), without relying on radial velocities, as such data were unavailable for the full photometric dataset \citep[see][]{paper1}. The uncertainties in the derived orbital parameters were estimated via a Monte Carlo simulation, where each orbit was integrated 500 times with input values perturbed according to the astrometric measurement uncertainties. For the stars of interest near $L_Z \sim 0~\kks$, the median uncertainties are $\sigma(r_{\rm ap}) = 0.1$~kpc, $\sigma(\zmax) = 0.2$~kpc, $\sigma(e) = 0.02$, and $\sigma(L_Z) = 45\ \kks$.

For all HPMS samples in this study, we required $\zmax$ to be greater than $2$~kpc in order to reduce contamination from disc stars. The main data sets used in the following analysis comprise $9012$, $9902$, $2778$, and $1333$ stars from SDSS, LAMOST, GALAH, and APOGEE, respectively, together with $11506$ stars from the photometric sample.

\subsection{Biases and systematics}\label{sec:zp}

Various surveys often adopt complex and distinct target-selection strategies, making direct comparisons between them nontrivial. Nonetheless, the appearance of the HPMS across datasets, as shown in Fig.~\ref{fig:hpms}, provides a practical diagnostic for assessing the relative biases of each survey. Compared to the HPMS identified in the photometric sample, the SDSS sample is strongly biased towards GSE stars with lower metallicity and smaller $L_Z$. In contrast, both GALAH and APOGEE include more metal-rich stars, which better delineate the vertical segment of the HPMS, while showing a significantly weaker presence of the GSE component.

Among the spectroscopic datasets, the LAMOST sample exhibits a shape and distribution most similar to the photometric HPMS (panel~(a)). Moreover, its overall extent and peak density in phase space are comparable to those of the photometric sample. While photometric catalogues generally contain significantly more stars than spectroscopic ones, this apparent similarity arises from specific limitations of our photometric sample. First, we restricted the sample to stars located near the Galactic prime meridian in order to convert proper motion into $\vphi$ in the absence of radial velocity information. Second, our photometric metallicity estimator was calibrated only for main-sequence stars, thereby confining the sample to the local volume ($< 3$~kpc). In contrast, spectroscopic surveys such as LAMOST include giant stars at greater distances, enabling them to probe a more extended region of the MW with significantly larger samples.

Despite inherent sample biases, the spectroscopic samples shown in Fig.~\ref{fig:hpms} confirm that the HPMS is a genuine structure, rather than an artefact arising from systematic errors in photometric parameter estimates. Furthermore, unlike the photometric sample, the spectroscopic data enable a detailed examination of the elemental-abundance distributions of stars along the sequence. In this regard, our datasets contain a sufficient number of HPMS stars to provide a robust foundation for subsequent chemical analyses. Although selection effects in the spectroscopic samples may still influence the inferred chemical trends, the photometric data offer a valuable means of evaluating and correcting for these biases.

At [Fe/H] $>-1$, the HPMS appears as a nearly vertical structure in all samples, as shown in Fig.~\ref{fig:hpms}, although its exact location varies due to differences in the metallicity scales of each dataset. For the SDSS and LAMOST samples, the median [Fe/H] of stars with $1500 < L_Z < 1750\ \kks$, corresponding to the most prominent part of the vertical segment, is approximately $-0.70$, reflecting the use of the same analysis tool for both survey datasets. In contrast, the GALAH and APOGEE samples show the vertical arm shifted towards slightly higher metallicities, with [Fe/H] values of approximately $-0.50$ and $-0.44$, respectively. Notably, these values are consistent with that of the photometric HPMS ($-0.49$), lending support to the reliability of the photometric metallicity scale. Furthermore, all samples exhibit a nearly zero mean $L_Z$ for GSE stars at [Fe/H] $<-1$. This consistency particularly reinforces the reliability of our $L_Z$ estimates in the photometric sample, which were derived using proper motions alone.

\section{Tracing stars formed in a Galactic merger}

\subsection{Metallicity distributions as a function of $L_Z$}

\begin{figure*}
\centering
\begin{subfigure}{5.6cm}
    \centering
    \includegraphics[width=\textwidth]{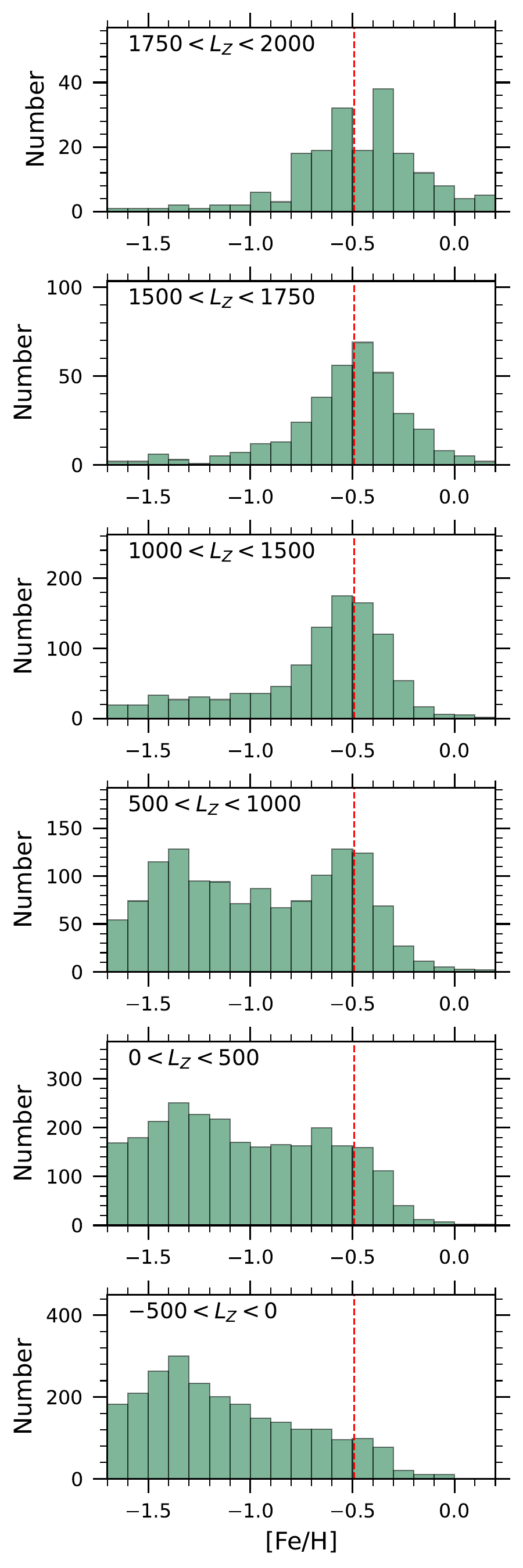}
        \caption{\textbf{Photometry}}
\end{subfigure}
\begin{subfigure}{5.6cm}
    \centering
    \includegraphics[width=\textwidth]{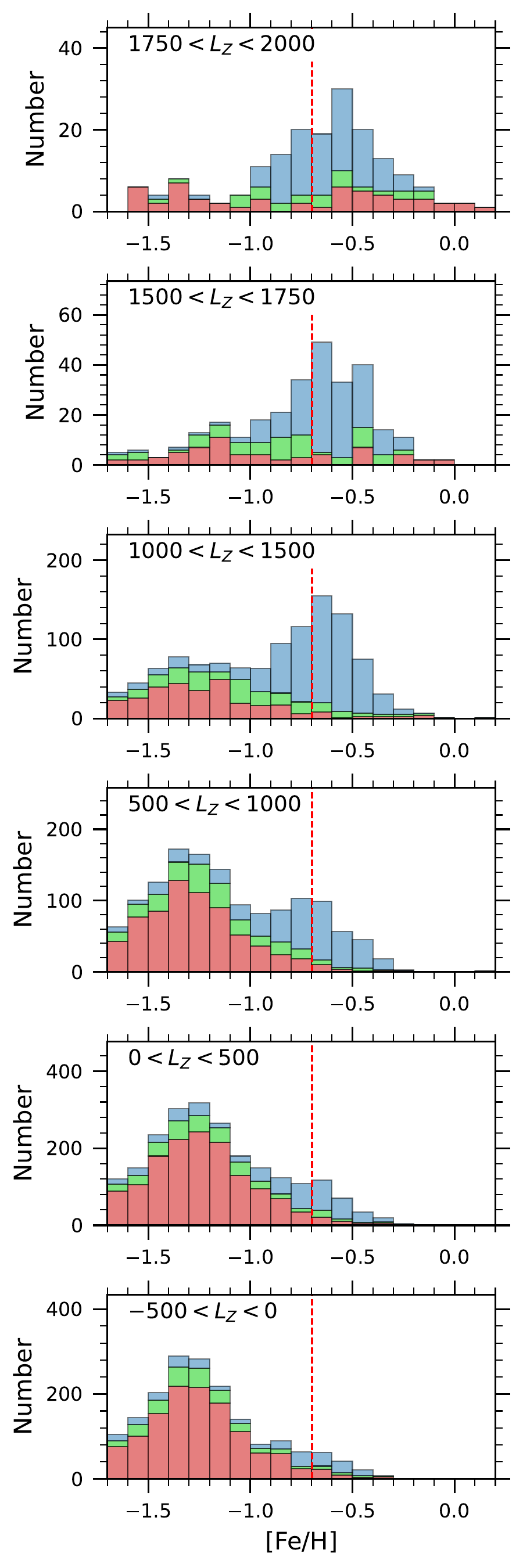}
        \caption{\textbf{SDSS}}
\end{subfigure}
\begin{subfigure}{5.6cm}
    \centering
    \includegraphics[width=\textwidth]{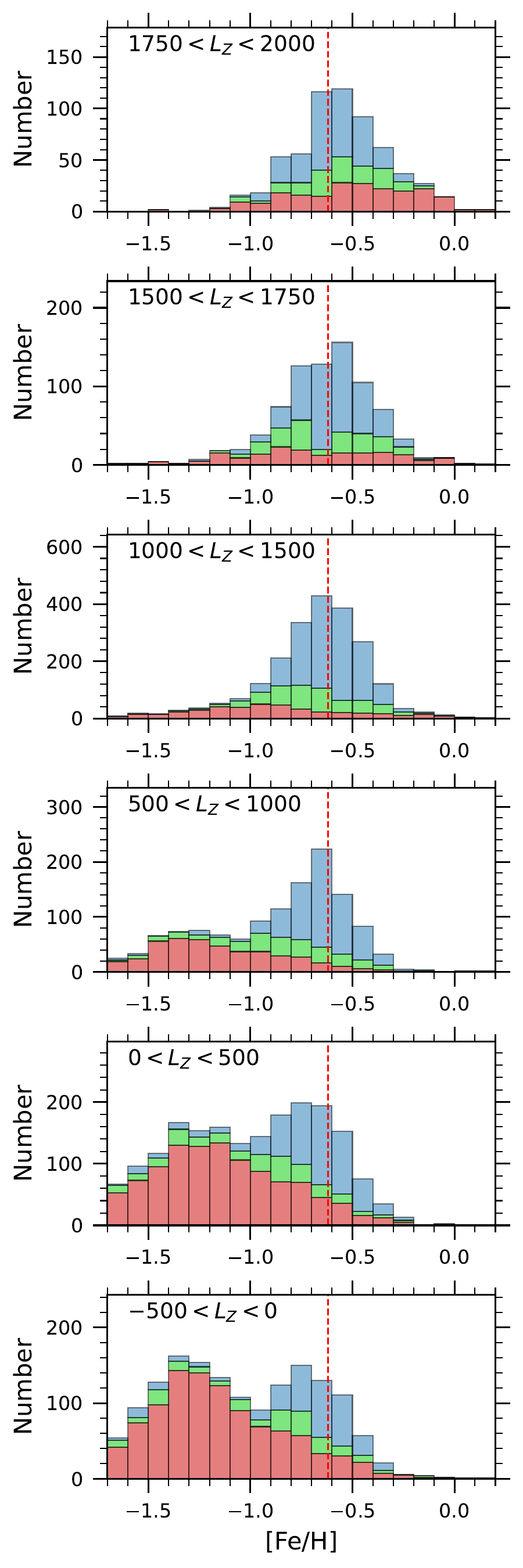}
    \caption{\textbf{LAMOST}}
\end{subfigure}
\caption{Metallicity distributions of the HPMS samples (Fig.~\ref{fig:hpms}) binned by $L_Z$. In all columns except Col.~(a), stars belonging to the low-$\alpha$ and high-$\alpha$ sequences are represented by red and blue histograms, respectively, while the green histogram indicates the uncertainty range due to variations in the $\alpha$-based division (see text for details). To guide the eye, vertical dashed lines mark the median [Fe/H] of all stars within $1500 < L_Z < 1750\ \kks$ in each sample.}\label{fig:mdf}
\end{figure*}

\begin{figure*}
\ContinuedFloat
\centering
\begin{subfigure}{5.6cm}
    \centering
    \includegraphics[width=\textwidth]{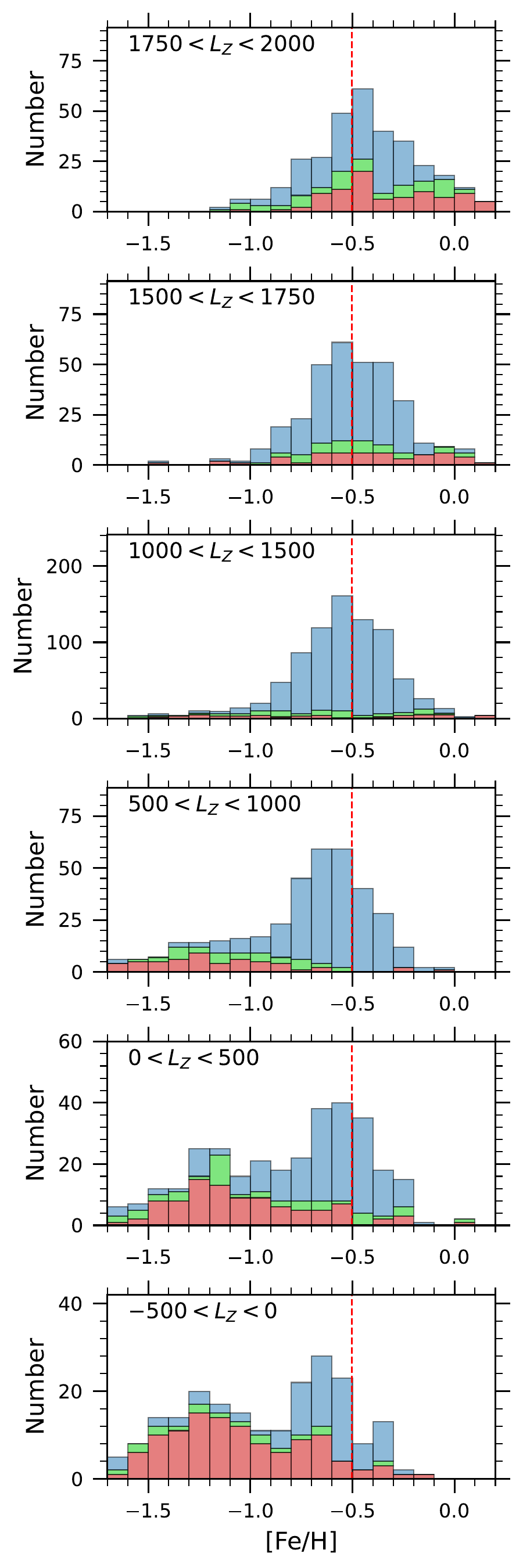}
        \caption{\textbf{GALAH}}
\end{subfigure}
\begin{subfigure}{5.6cm}
    \centering
    \includegraphics[width=\textwidth]{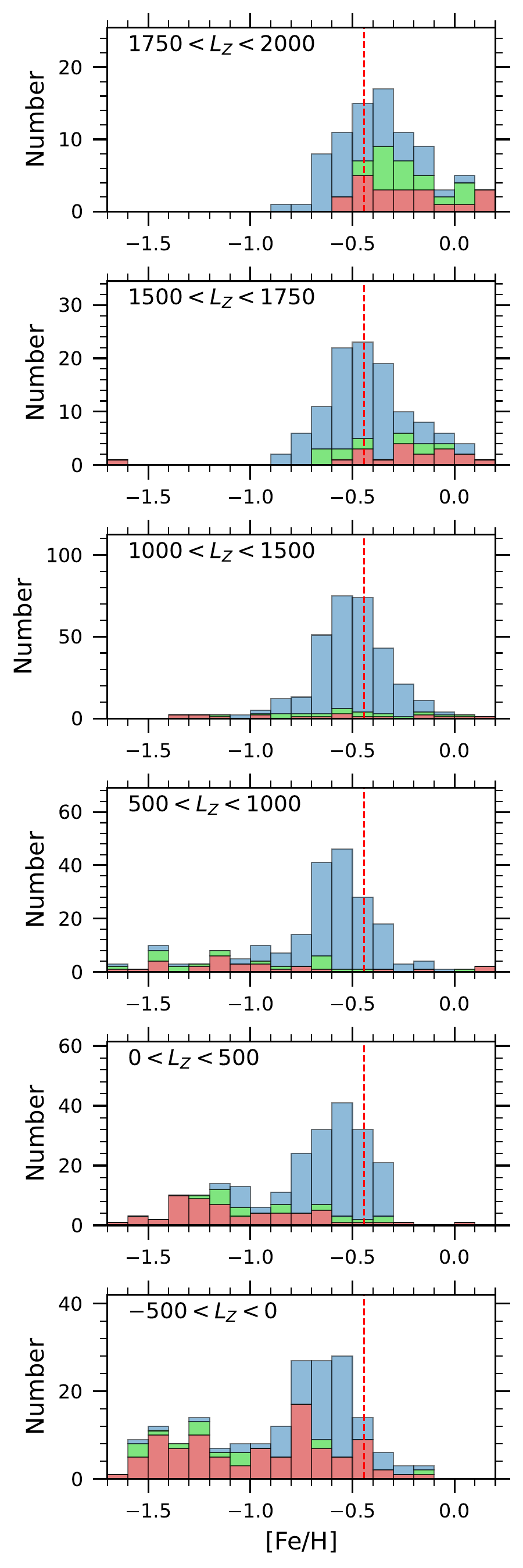}
    \caption{\textbf{APOGEE}}
\end{subfigure}
\caption{Continued. Cols.~(d) and (e) refer to the high-resolution samples from GALAH and APOGEE, respectively.}
\end{figure*}

Figure~\ref{fig:mdf} presents the metallicity distributions of our samples from photometry, SEGUE, LAMOST, GALAH, and APOGEE, respectively. Each dataset was binned by $L_Z$ to highlight the varying contributions of different stellar populations across the HPMS. In the spectroscopic samples, each bin was further divided into low-$\alpha$ (red histograms) and high-$\alpha$ (blue histograms) sequences (see Appendix~\ref{sec:appendix1}). For the GALAH and APOGEE samples, we adopted an uncertainty of $\Delta$[$\alpha$/Fe] $=\pm0.02$~dex for the division between the two $\alpha$ sequences, while a slightly larger uncertainty of $\Delta$[$\alpha$/Fe] $=\pm0.03$~dex was adopted for the medium-resolution spectroscopic samples. Green histograms indicate the level of uncertainty in the population fractions arising from variations in the $\alpha$-based division.

At high metallicity ([Fe/H] $> -1.0$), the spectroscopic samples show that stars on the high-$\alpha$ sequence (hereafter, high-$\alpha$ stars) clearly dominate over those on the low-$\alpha$ sequence (low-$\alpha$ stars) across most $L_Z$ bins. The peak metallicity, corresponding to the vertical part of the HPMS in Fig.~\ref{fig:hpms}, remains nearly constant at [Fe/H] $\approx -0.5$ to $-0.7$ over the range $-500 < L_Z < 1500\ \kks$, above which it tends to increase mildly. These high-$\alpha$ stars are associated either with the canonical thick disc or with stars formed through dynamical heating of the primordial disc. The photometric sample in the left-most column also shows a peak at [Fe/H] $\sim -0.5$; however, this metallicity peak becomes less prominent at low $L_Z$ ($<500\ \kks$), indicating that the metal-rich component has a weaker influence than in the spectroscopic samples. Such systematic differences likely arise from sample biases.

In contrast, the low-metallicity peak at [Fe/H] $=-1.3$ to $-1.4$, seen in all samples, is dominated by low-$\alpha$ stars, as is evident from the spectroscopic samples. These low-$\alpha$ stars are predominantly found in the low-$L_Z$ regime ($-500 \la L_Z \la 500\ \kks$), consistent with the interpretation that they are mostly debris from the GSE progenitor galaxy. The contribution of low-$\alpha$ stars declines rapidly at $L_Z > 1000\ \kks$, but they reappear at larger $L_Z$, forming a rotationally supported component. These stars, however, exhibit significantly higher metallicities of [Fe/H] $> -0.7$ and are more closely related to the canonical thin disc, despite the $\zmax > 2$~kpc cut applied to the sample.

\begin{figure}
\centering
\includegraphics[width=8.2cm]{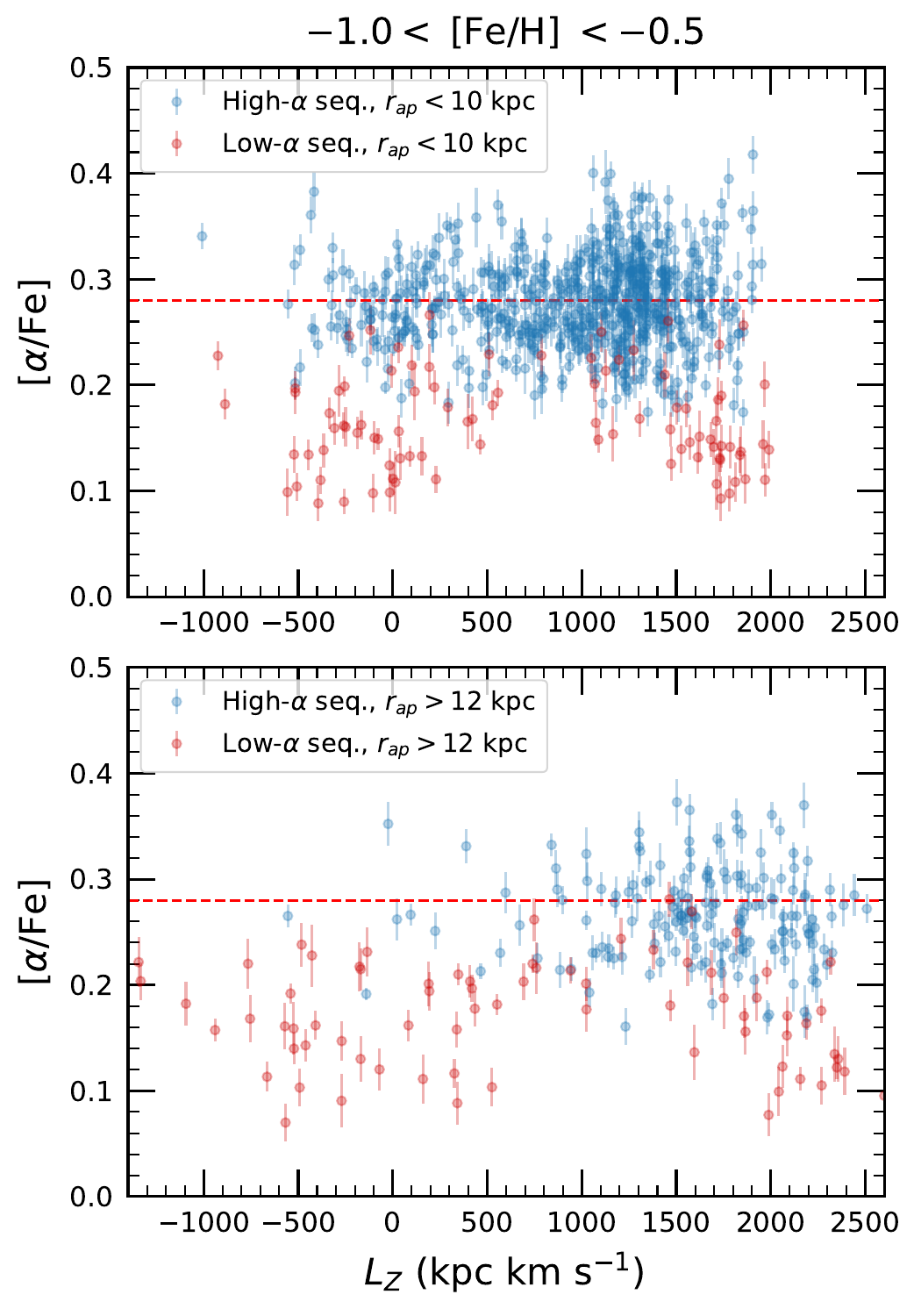}
\caption{Distribution of [$\alpha$/Fe] for GALAH stars in the vertical segment of the HPMS ($-1.0 < {\rm [Fe/H]} < -0.5$). The top and bottom panels show [$\alpha$/Fe] as a function of $L_Z$ for stars in the inner ($\rap < 10$~kpc) and outer ($\rap > 12$~kpc) regions, respectively. High-$\alpha$ stars are shown in blue and low-$\alpha$ stars in red. The high-$\alpha$ population dominates the inner region and exhibits a relatively constant [$\alpha$/Fe] over a wide range of $L_Z$. In contrast, the low-$\alpha$ stars display a bimodal $L_Z$ distribution: the low-$L_Z$ stars are primarily associated with GSE, while the high-$L_Z$ stars may trace the high-$\zmax$ tail of the thin-disc population.}
\label{fig:galah_lz}
\end{figure}

The bimodal distribution of $L_Z$ among low-$\alpha$ stars is also evident in Fig.~\ref{fig:galah_lz}, which displays the distribution of $\afe$ as a function of $L_Z$ along the vertical arm of the HPMS ($-1.0 < {\rm [Fe/H]} < -0.5$). For stars whose orbits are primarily confined to the inner region ($\rap < 10$~kpc; top panel), the low-$\alpha$ stars exhibit two distinct peaks at $L_Z \sim 0$ and $1800\ \kks$, corresponding to the GSE and disc populations, respectively. A similar bimodal pattern is observed among low-$\alpha$ stars in the outer region ($\rap > 12$~kpc; bottom panel), indicating that these populations extend far into the outskirts of the MW.

On the contrary, high-$\alpha$ stars with small $\rap$ (blue circles in the top panel of Fig.~\ref{fig:galah_lz}) maintain a nearly constant $\afe$ across a wide range of $L_Z$. Notably, this sequence extends to negative rotation ($L_Z \approx -500~\kks$), indicating the presence of a dynamically heated population originating from a chemically well-mixed primordial disc. Meanwhile, the outer region (bottom panel), which consists of stars with large $\rap$, does not prominently exhibit the Splash population. This suggests that the heated-disc population is primarily confined to the inner MW, consistent with a scenario in which the primordial disc was relatively compact when it was disrupted by a merger \citep[e.g.][]{belokurov:20,paper3}.  

Based on an inspection of the [Fe/H], [$\alpha$/Fe], and $L_Z$ distributions in Figs.~\ref{fig:mdf} and \ref{fig:galah_lz}, it is evident that stars along the HPMS comprise at least three distinct and well-characterised populations:

\begin{itemize}

\item GSE debris: These metal-poor stars (primarily with ${\rm [Fe/H]} \la -1.0$) follow the low-$\alpha$ sequence and are characterised by low $L_Z$ and negligible net rotation.

\item High-$\alpha$ population: These metal-rich stars ($-1.0 \la {\rm [Fe/H]} \la -0.4$) span $L_Z$ values from approximately $-500$ to $2000\ \kks$. This group includes thick-disc stars as well as a dynamically heated population (often referred to as Splash), both likely sharing a common chemical origin.

\item Metal-rich, low-$\alpha$ stars: With ${\rm [Fe/H]} \ga -0.7$ and high angular momentum ($L_Z \ga 1500\ \kks$), these low-$\alpha$ stars may be part of the vertically extended tail of the canonical thin-disc population.

\end{itemize}

In addition to these populations, the spectroscopic metallicity distributions from GALAH (Col.~(d) in Fig.~\ref{fig:mdf}) clearly reveal the presence of a fourth population. In the lower two panels ($-500 < L_Z < 500\ \text{kpc km s}^{-1}$), the metallicity distribution of low-$\alpha$ stars (red histograms) displays an excess of metal-rich stars, with peaks at nearly the same metallicity as the heated-disc population ([Fe/H] $\approx -0.65$). Their low $\alpha$-element abundances, combined with relatively high metallicities, suggest a possible connection to thin-disc stars. However, their low $L_Z$ values rule out the possibility that they are low-$\alpha$ counterparts of the heated-disc population. Moreover, while high-$\alpha$ stars consistently exhibit elevated $\alpha$-element abundances across the full $L_Z$ range, the low-$\alpha$ stars display a bimodal $L_Z$ distribution (Fig.~\ref{fig:galah_lz}). Rather than being associated with the heated disc, the combination of low $\alpha$-element abundances and low $L_Z$ instead points to a closer connection with the GSE population. To further investigate their origins, we examined high-resolution spectroscopic abundances from GALAH and APOGEE in the following section, which reveal distinct chemical signatures that clearly distinguish them from the accreted GSE population.

\subsection{Detailed chemical signatures}\label{sec:LAHN}

\begin{figure*}
\centering
\includegraphics[width=16cm]{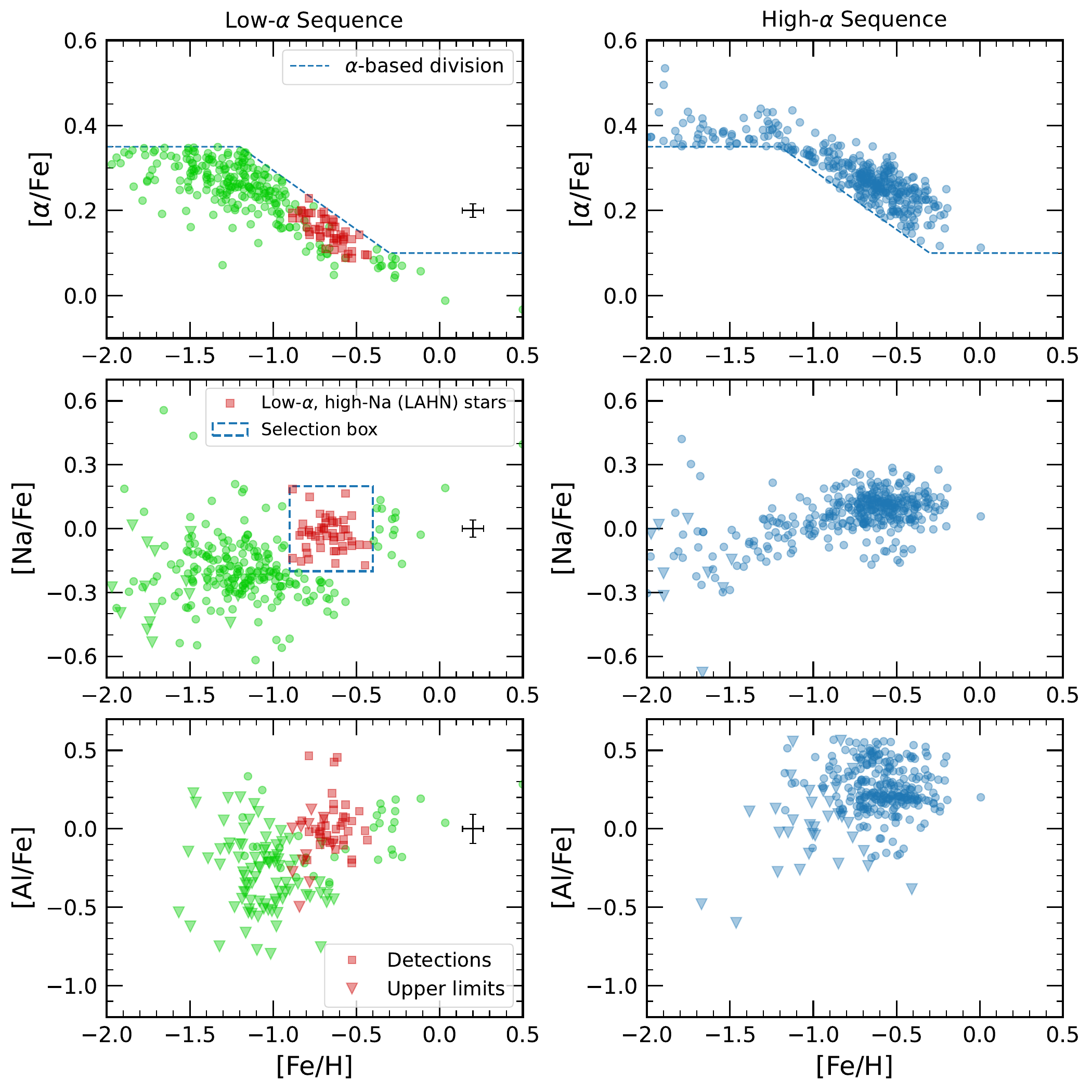}
\caption{Chemical abundances ([$\alpha$/Fe], [Na/Fe], and [Al/Fe]) of the GALAH sample with $|L_Z| < 600~\kks$. In the top panels, the dashed blue line indicates the division based on [$\alpha$/Fe], separating stars into low-$\alpha$ (left) and high-$\alpha$ (right) sequences. Red squares and triangles in the left column mark low-$\alpha$, high-Na (LAHN) stars. The dashed blue box in the middle left panel outlines this selection. A representative median error is shown.}
\label{fig:abun}
\end{figure*}

The first line of evidence suggesting that the excess of high-[Fe/H], low-$\alpha$ stars has an origin distinct from the directly accreted GSE population comes from detailed elemental abundance ratios in the GALAH and APOGEE samples. This is demonstrated clearly in Fig.~\ref{fig:abun}, which presents [$\alpha$/Fe], [Na/Fe], and [Al/Fe] as functions of [Fe/H] from GALAH. In this analysis, we included stars with upper measurement limits (shown as triangles), but excluded those with flagged abundance values ({\tt flag\_X\_fe}). All panels show stars with $|L_Z| < 600\ \kks$, where the excess of low-$\alpha$ stars is most prominent in the metallicity distributions (Fig.~\ref{fig:mdf}).

The top panels of Fig.~\ref{fig:abun} divide the sample into low-$\alpha$ (left) and high-$\alpha$ (right) stars. Each sequence follows a relatively tight [$\alpha$/Fe]--[Fe/H] relation and remains largely distinct from the other. The GSE debris is also well represented in this $L_Z$ range, as indicated by the prominent low-$\alpha$ sequence. However, as is shown in the middle left panel, the low-$\alpha$ stars branch into two groups in the [Fe/H]--[Na/Fe] plane at [Fe/H] $\ga -0.9$. The main body of GSE stars has [Na/Fe] $\approx -0.2$ at [Fe/H] $\sim -1.3$ and trends towards lower [Na/Fe] with increasing metallicity, reaching [Na/Fe] $\approx -0.3$ at [Fe/H] $\sim -0.7$. In contrast, at $-0.9 < {\rm [Fe/H]} < -0.4$, where the excess of low-$\alpha$ stars appears in the metallicity distributions, more than half of them (48 out of 69) exhibit elevated Na abundances around [Na/Fe] $\sim0$.

The red squares in the left panels of Fig.~\ref{fig:abun} highlight these low-$\alpha$, high-Na (LAHN) stars, which are defined as those lying within the dashed blue box in the middle left panel. While some stars at higher [Fe/H], or even some high-$\alpha$ stars with systematically lower Na abundances, may be included in this selection, the [Fe/H]--[Na/Fe] cut is designed to isolate the LAHN stars and enable comparisons with the low-$\alpha$, Na-poor GSE stars in the same metallicity range. Under this selection, the LAHN stars in the top left panel exhibit a subtle but systematic shift towards higher [$\alpha$/Fe] relative to their Na-poor counterparts. A similar trend is seen in [Al/Fe] in the bottom left panel: although Al detections in GALAH are more challenging, with many measurements reported as upper limits, the available data reveal a population of low-$\alpha$ stars with enhanced Al abundances compared to other low-$\alpha$ stars at similar [Fe/H].

\begin{figure*}
\centering
\includegraphics[width=16cm]{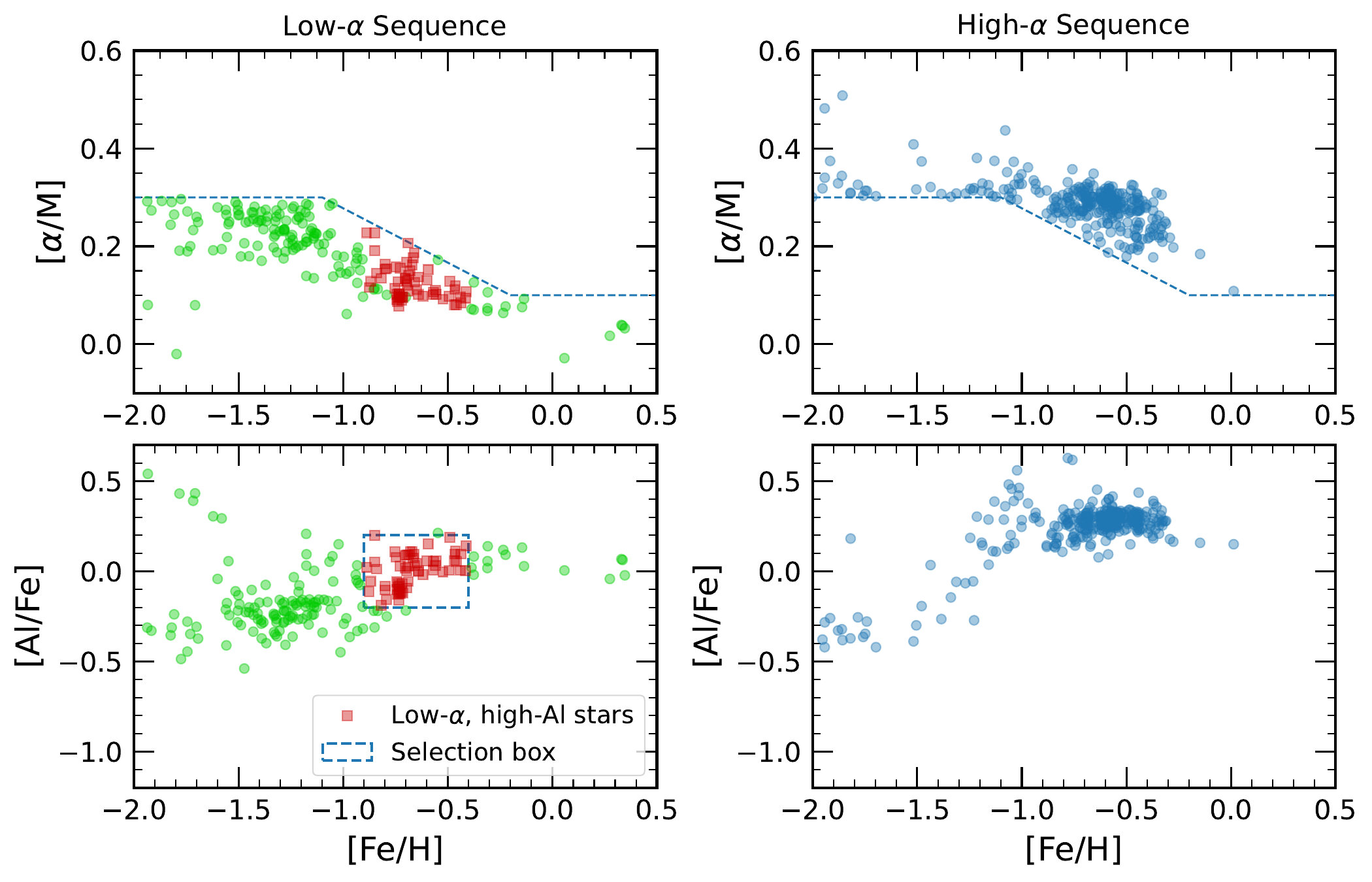}
\caption{Same as Fig.~\ref{fig:abun}, but displaying [$\alpha$/M] and [Al/Fe] from the APOGEE dataset for stars with $|L_Z| < 600\ \kks$. The dashed blue box indicates the region in the [Fe/H]--[Al/Fe] plane where LAHN stars were identified in the GALAH sample (bottom left panel of Fig.~\ref{fig:abun}).}
\label{fig:abun2}
\end{figure*}

A weaker but qualitatively similar chemical trend is seen in the APOGEE sample (Fig.~\ref{fig:abun2}). Since APOGEE DR17 provides [Al/Fe] measurements but lacks [Na/Fe], we utilised [$\alpha$/M] and the Al abundances. To guide our interpretation, we used the Al abundance range of the LAHN stars in the GALAH sample, as indicated by the dashed blue box in the bottom left panel. This selection is approximate and intended only as a reference. The stars falling within this selection box are highlighted by red squares. Although there is substantial overlap between LAHN stars and GSE debris, there is a modest but consistent indication that the LAHN stars exhibit systematically higher Al and $\alpha$ abundances compared to those likely associated with GSE.

Light elements such as Na and Al can be produced in asymptotic giant branch stars, but their primary synthesis occurs in core-collapse supernovae (SNe), where they are generated in greater quantities during the explosive nucleosynthesis of massive stars \citep[see][and references therein]{johnson:19}. Because core-collapse SNe are closely linked to active star formation, Na, Al, and $\alpha$-elements serve as valuable tracers of star-formation history and chemical evolution. In this context, the presence of chemical bimodality in Na and Al among low-$\alpha$ stars, along with a weaker bimodality in the $\alpha$ elements, provides compelling evidence for two distinct chemical-enrichment pathways. The low-$\alpha$, low-Na group corresponds to accreted stars from GSE that formed within its progenitor dwarf galaxy, where chemical enrichment was slower and less efficient. In contrast, the LAHN group likely formed during episodes of intense star formation, as indicated by their systematically elevated Na and Al abundances, along with slightly enhanced $\alpha$-element levels compared to the GSE stars. These chemical signatures point to a stronger contribution from core-collapse SNe to their enrichment history.

The LAHN population shows a close connection to the intermediate [$\alpha$/Fe] stars identified by \citet[][termed `Eos']{myeong:22}. They selected stars with highly eccentric orbits ($e > 0.85$) from APOGEE DR17 and GALAH DR3, and identified a group with a mean [Fe/H] $\approx -0.7$ and [$\alpha$/Fe] $\approx +0.1$ using a Gaussian mixture model. Despite considerable scatter, these stars occupy a region between the GSE and heated-disc populations in the [Fe/H]--[$\alpha$/Fe] plane, consistent with our findings. Moreover, their mean elemental abundances of [Na/Fe] $\approx-0.1$ and [Al/Fe] $\approx0.0$ strongly resemble ours. Furthermore, Eos stars fall within a similar angular momentum range \citep[$|L_Z| \la 500\ \kks$;][]{matsuno:24}, indicating that a fair fraction of our LAHN population belongs to the Eos group. Despite this notable overlap, our identification of the LAHN population builds on the characterisation of the HPMS \citep{paper4}, a previously unrecognised stellar sequence linked to the GSE merger. A more detailed discussion of the relationship between the LAHN and Eos populations will be presented in \S~\ref{sec:discussion}, following an examination of the spatial and orbital properties of the LAHN stars.

\subsection{Spatial and orbital characteristics}

\begin{figure*}
\centering
\includegraphics[width=14cm]{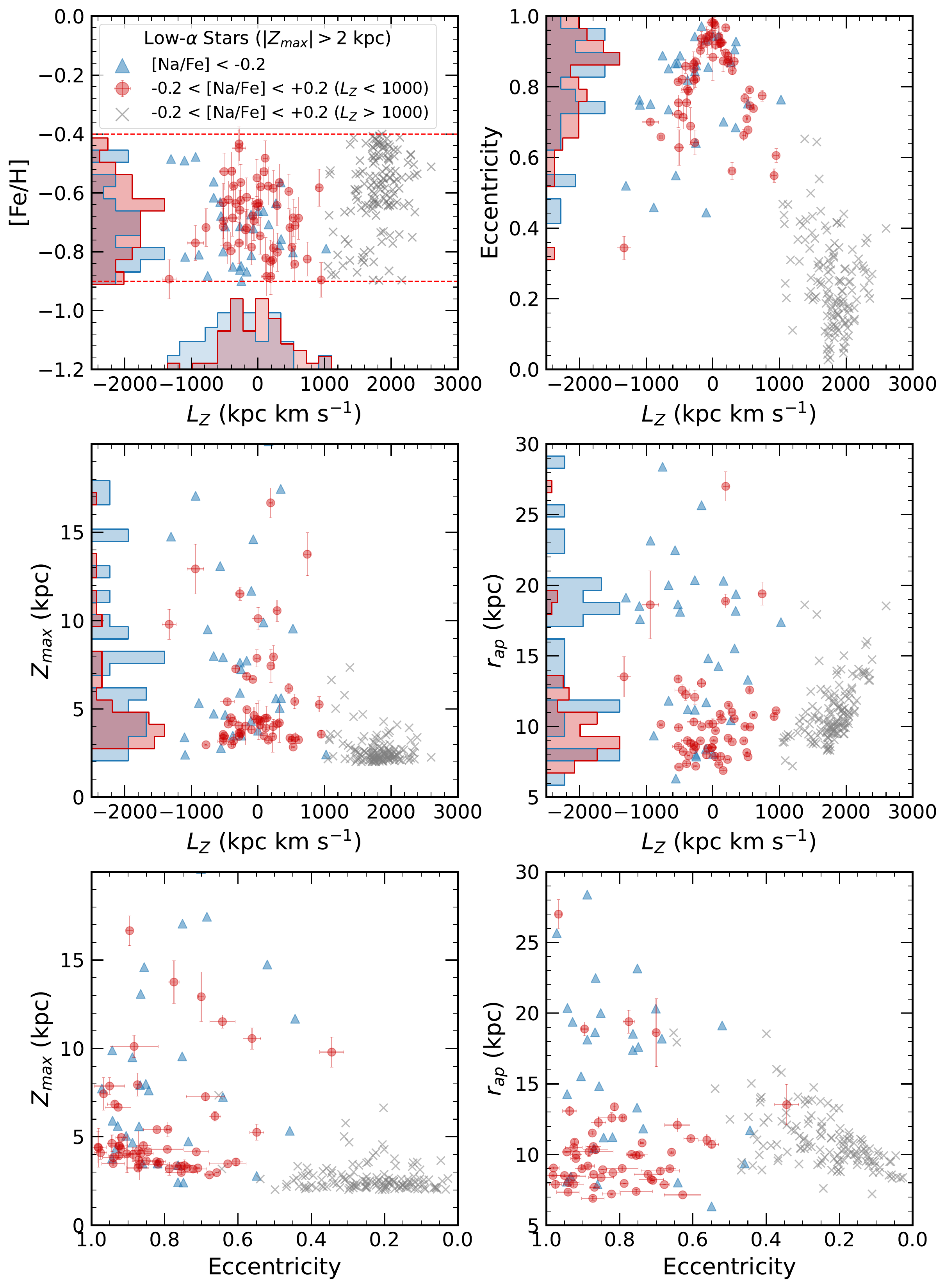}
\caption{Spatial and orbital properties of low-$\alpha$ stars with $-0.9 < {\rm [Fe/H]} < -0.4$ in the GALAH sample. The stars were divided into two groups based on their [Na/Fe] abundances (see the middle left panel of Fig.~\ref{fig:abun}): the LAHN stars are displayed as red circles, while other low-$\alpha$ stars with lower [Na/Fe], likely associated with the accreted GSE debris, are indicated by blue triangles. Error bars are shown only for the LAHN group. Histograms display the distributions of each group using the same colour scheme. In addition to these two groups, high-$L_Z$ ($> 1000~\kks$; disc-like) stars that meet the same abundance criteria as the LAHN population are marked with grey crosses. Corresponding plots for high-$\alpha$ stars are provided in Appendix~\ref{sec:appendix2}.}
\label{fig:asp}
\end{figure*}

To further investigate the nature and origin of the LAHN stars, we examined their orbital properties within the MW. Figure~\ref{fig:asp} shows the distributions of $L_Z$, $\zmax$, orbital eccentricity, and apogalacticon for low-$\alpha$ stars in the GALAH sample, selected along the vertical arm of the HPMS ($-0.9 < {\rm [Fe/H]} < -0.4$). The sample is divided into two groups based on their [Na/Fe] abundances: the LAHN stars ($-0.2 < {\rm [Na/Fe]} < +0.2$; red circles) and the GSE population with [Na/Fe] $< -0.2$ (blue triangles). Histograms along each axis show the distributions of these two groups separately. As is discussed above (Fig.~\ref{fig:galah_lz}), low-$\alpha$ stars exhibit a bimodal $L_Z$ distribution, with peaks at $L_Z \sim 0$ and $1800\ \kks$, separated by a sparsely populated region. To distinguish stars with GSE-like kinematics, we adopted a division at $L_Z = 1000\ \kks$. Stars that meet the same selection criteria as the LAHN group but lie above this $L_Z$ threshold are shown as grey crosses. Additional plots for the high-$\alpha$ population are provided in Appendix~\ref{sec:appendix2}.

Three key features of the LAHN stars are evident in Fig.~\ref{fig:asp}:

\begin{itemize}

\item They have high orbital eccentricities, comparable to those of accreted GSE stars (i.e.\ low-Na stars).

\item They are more tightly confined to the inner MW compared to the accreted component. This is evident in their maximum vertical excursions: about $75\%$ of the LAHN stars have $\zmax < 6$~kpc, whereas the corresponding fraction for the accreted stars is only $50\%$. Most LAHN stars also have apogalacticons within $\rap < 12$~kpc, while the low-Na group is more extended, with a possible accumulation of GSE stars near $\rap \sim 20$~kpc \citep[see][]{deason:18,naidu:21}.

\item Among the LAHN stars, those with higher orbital eccentricities tend to reach higher $\zmax$. This coupling between $\zmax$ and eccentricity reflects a shared dynamical origin of the LAHN stars and the dynamically heated population, likely tracing back to the same merger event \citep[see also][]{khoperskov:23}.

\end{itemize}

Since the orbital characteristics (eccentricities and $L_Z$) of the LAHN stars closely resemble those of stars accreted from GSE, it follows that the clouds in which the LAHN stars formed originally belonged to the progenitor galaxy of GSE rather than the primordial MW. Their closer alignment with the low-$\alpha$ sequence further supports this origin, indicating that these clouds were part of the dwarf galaxy. As star-forming gas clouds from the GSE progenitor lost angular momentum during the merger, they may have been funnelled into the inner MW, leading to the formation of LAHN stars with more compact orbits and tighter confinement to the disc plane than stars directly accreted from GSE. Therefore, these spatial and orbital signatures provide a second, independent line of evidence supporting their merger-driven origin.

In addition, Fig.~\ref{fig:asp} hints at a possible connection between the LAHN population and the canonical thin disc. The metallicity distributions of low-$\alpha$, high-$L_Z$ disc stars (Fig.~\ref{fig:mdf}) reveal a low-metallicity cut-off at [Fe/H] $\approx-0.7$, coinciding with the metallicity range where LAHN stars are predominantly found. Furthermore, the LAHN stars exhibit near-solar Na and Al abundances -- higher than those of GSE but lower than those of the canonical thick disc or heated-disc population. As is shown by the grey crosses in Fig.~\ref{fig:asp}, these stars have orbits that are more confined to the disc, yet are more radially extended than those of the LAHN stars, consistent with an upside-down, inside-out formation of the Galactic thin disc \citep[e.g.][]{bird:13}. If thin-disc stars had formed exclusively from gas clouds in the primordial MW, such a coincidence in chemical and dynamical properties would be less likely.

\subsection{Evidence from numerical simulations}

\begin{figure*}
\centering
\begin{subfigure}{6.5cm}
    \centering
    \includegraphics[width=\textwidth]{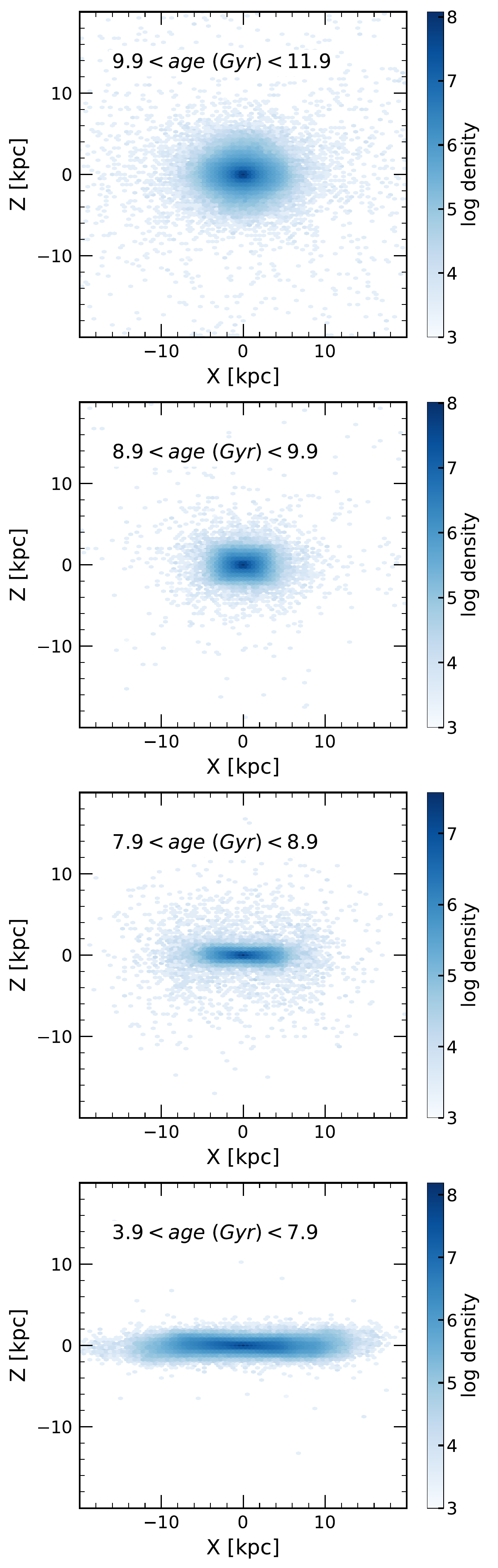}
            \caption{\textbf{Spatial distribution}}
\end{subfigure}
\hspace{1cm}
\begin{subfigure}{7cm}
    \centering
    \includegraphics[width=\textwidth]{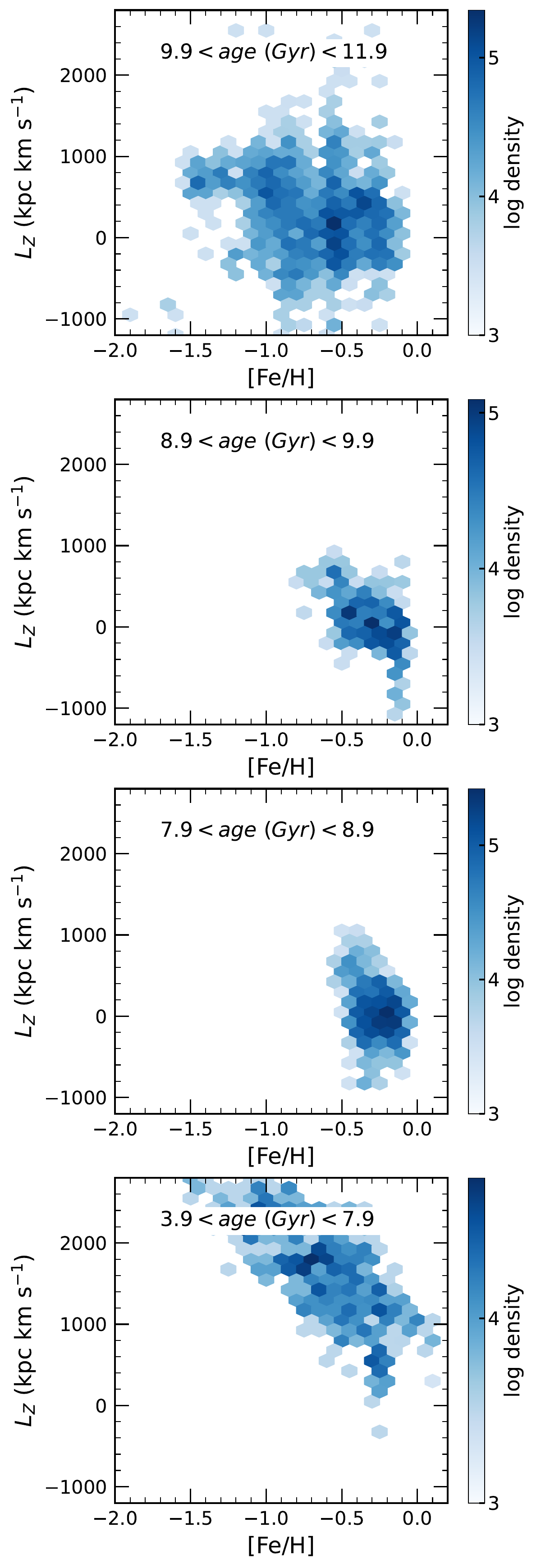}
        \caption{\textbf{[Fe/H] vs.\ $L_Z$}}
\end{subfigure}
\caption{Cosmological zoom-in simulation of a MW-like galaxy. Only stars that formed within the main halo are shown, excluding those accreted from dwarf galaxies. The left column displays the spatial distributions in an edge-on X-Z projection at $z = 0$, divided into four age bins. The right column shows the corresponding phase-space distributions of star particles in the [Fe/H]--$L_Z$ space, selected from a region near the solar circle ($6 < R < 20$~kpc and $2 < |Z| < 10$~kpc). Star particles with ages between 7.9 and 9.9 billion years (middle panels) originated during a burst of star formation following the major merger at $z \sim 2$. Stars formed before this period (top panels) exhibit characteristics similar to the heated-disc population in the MW, whereas younger stars (bottom panels) contributed to the disc’s inside-out growth.}
\label{fig:sim}
\end{figure*}

The final piece of evidence supporting the interpretation of LAHN stars as a distinct population formed during an episode of intense star formation comes from a cosmological zoom-in simulation of a MW-like galaxy \citep{hirai:22}, conducted using the $N$-body/smoothed particle hydrodynamics code ASURA \citep{saitoh:08, saitoh:09}. The simulation follows the chemo-dynamical evolution of satellite galaxies within a MW-mass host, with a virial mass of $1.2 \times 10^{12}\ M_\odot$, making it well suited for studying the origin of stellar populations contributed by accreted dwarf galaxies. It includes key physical processes, such as metallicity-dependent radiative cooling and heating \citep{ferland:13}, star formation based on the \citet{schmidt:59} law with realistic thresholds \citep{hirai:21}, and chemical enrichment from core-collapse SNe, Type Ia SNe, and neutron star mergers \citep{saitoh:17}. Feedback mechanisms, such as Lyman-$\alpha$ heating from massive stars \citep{fujii:21} and turbulent metal diffusion \citep{hirai:17}, are also incorporated, allowing for realistic self-regulation of star formation. Star particles were modelled as simple stellar populations with the \citet{chabrier:03} initial mass function.

Figure~\ref{fig:sim} presents the spatial (left) and [Fe/H]--$L_Z$ (right) distributions of star particles in the simulation at $z = 0$. Since our primary goal is to identify stars formed during merger-driven star formation, we show only those that formed within the main halo (i.e.\ in situ stars). To generate mock results comparable to our observational data, we applied a spatial cut of $6 < R < 20$~kpc and $2 < |Z| < 10$~kpc, selecting stars within a region that encompasses the solar annulus. Unlike our observational samples, no additional selection based on proper motions was applied. In Fig.~\ref{fig:sim}, the panels are divided into four age bins based on major transition periods in the galaxy's evolution: (1) stars that formed prior to the last major merger ($z \sim 2$; top row), (2) stars that formed during the bursty and episodic star formation between $z \approx 1.7$ and $z \approx 1.1$ (middle two rows), and (3) stars that formed after the period of intense star formation (bottom row).

By the time the last major merger occurred, approximately half of the halo mass at $z = 0$ had been assembled. As is shown in the top row of Fig.~\ref{fig:sim}, most stars formed prior to the significant merger events originated in the primordial disc and were subsequently scattered into the halo, marking them as a heated-disc population. Their quasi-spherical spatial distribution is consistent with that observed in the MW \citep{belokurov:20,paper3}. Their metallicity distribution ($-1.2 \la {\rm [Fe/H]} \la -0.2$), together with their broad range in $L_Z$ ($-500 \la L_Z \la 1500\ \kks$), further aligns with the chemo-dynamical signatures of dynamically heated stars (see Fig.~\ref{fig:hpms}).

The middle two rows of Fig.~\ref{fig:sim} present the phase-space distributions of stars with ages of $7.9$--$9.9$ billion years, subdivided into two age bins. These stars formed during bursty episodes of intense star formation triggered by the merger and supernova feedback, with successive peaks separated by a few hundred million years, highlighting the imprint of the GSE-like merger event. They are confined to a relatively narrow metallicity range ($-0.6 < {\rm [Fe/H]} < -0.1$) and exhibit low $L_Z$ values ($|L_Z| \la 1000\ \kks$), consistent with our findings for the LAHN population. Moreover, their spatial distribution at $z = 0$ is more compact than that of older stars and becomes increasingly flattened towards the Galactic plane for the slightly younger population. This spatial behaviour is also characteristic of the LAHN stars, which show a more concentrated distribution compared to accreted GSE stars. Interestingly, the phase-space distribution in [Fe/H]--$L_Z$ remains nearly unchanged between the two age bins. This suggests that the star-formation process was highly chaotic, rather than one in which star formation occurred as gas clouds gradually settled into the disc.

Unlike in the observational samples, we did not classify star particles in the simulation based on their $\alpha$-element abundances. The primary purpose of our $\alpha$-based division was to distinguish heated-disc stars; however, such a division is unnecessary in the simulation, where heated-disc populations can be directly identified based on stellar ages (top panels in Fig.~\ref{fig:sim}). Nonetheless, elemental abundances in numerical simulations provide valuable constraints on the properties of starburst populations, and will be explored in greater detail in a forthcoming study (Hirai et al., in prep.). Briefly, our simulation, which includes a detailed chemical evolution model, shows that stars formed in a merger-driven starburst exhibit significant chemical inhomogeneities in both $\alpha$- and light elements. This suggests that some stars formed directly from the ejecta of core-collapse SNe, while others were primarily enriched by Type Ia SNe from earlier generations.

Following the period of intense star formation, the simulated galaxy experienced more continuous star formation, building up the disc in an inside-out fashion, as evident in the bottom-left panel of Fig.~\ref{fig:sim}. Our simulation indicates that young disc stars formed from gas enriched by inflowing clouds associated with merger events, supporting the idea that the material that gave rise to the LAHN stars may have also contributed to the formation of the MW's thin disc. Interestingly, these young disc stars exhibit a negative correlation between $L_Z$ and [Fe/H], mirroring the anti-correlation observed in the Galactic thin disc \citep[e.g.][]{lee:11}. Nonetheless, our observed HPMS sample does not show such a relation, but appears clustered in phase space at the high-$\vphi$ end. This is due to the stringent proper-motion constraints imposed on the sample, which excluded high-metallicity thin-disc stars with $\vphi$ values similar to that of the Sun.

We also note that our simulation does not produce a well-defined thick-disc structure, which would otherwise show a positive correlation between [Fe/H] and $L_Z$. This may be due to the selection criteria used for the simulated galaxy, which were based on mass and assembly history, rather than specific properties such as the presence of a thick disc. Conversely, it could be that stars in the primordial-disc component were fully dispersed into the halo by dynamical heating, leaving behind no coherent thick-disc population distinguishable in the chemo-dynamical space.

\section{Relative fraction of the LAHN population}

\begin{figure*}
\centering
\begin{subfigure}{8.4cm}
    \centering
    \includegraphics[width=\textwidth]{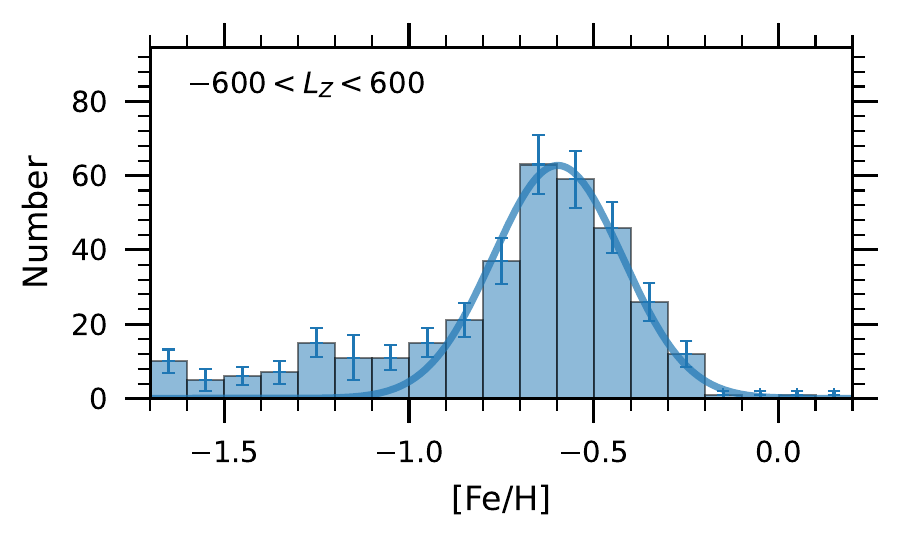}
    \caption{\textbf{GALAH}}
\end{subfigure}
\vspace{0.4cm}
\begin{subfigure}{8.4cm}
    \centering
    \includegraphics[width=\textwidth]{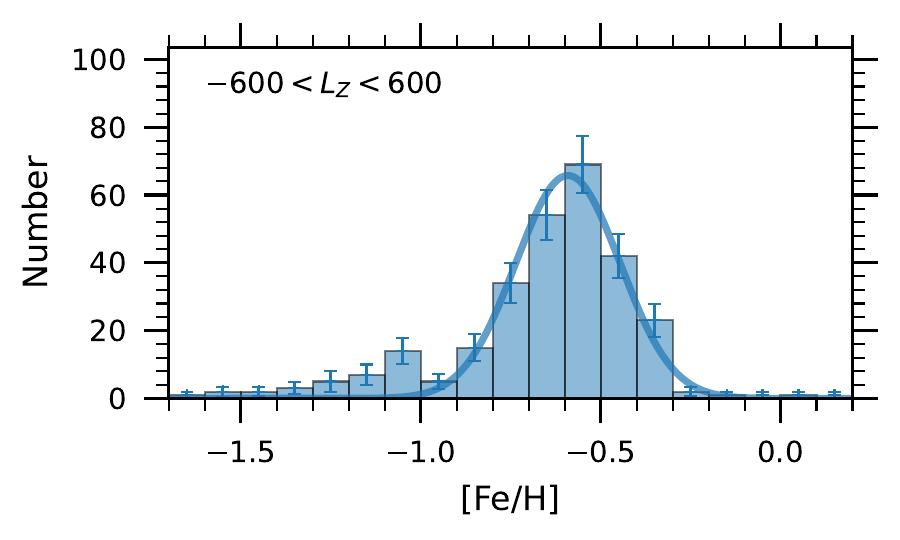}
    \caption{\textbf{APOGEE}}
\end{subfigure}
\begin{subfigure}{8.4cm}
    \centering
    \includegraphics[width=\textwidth]{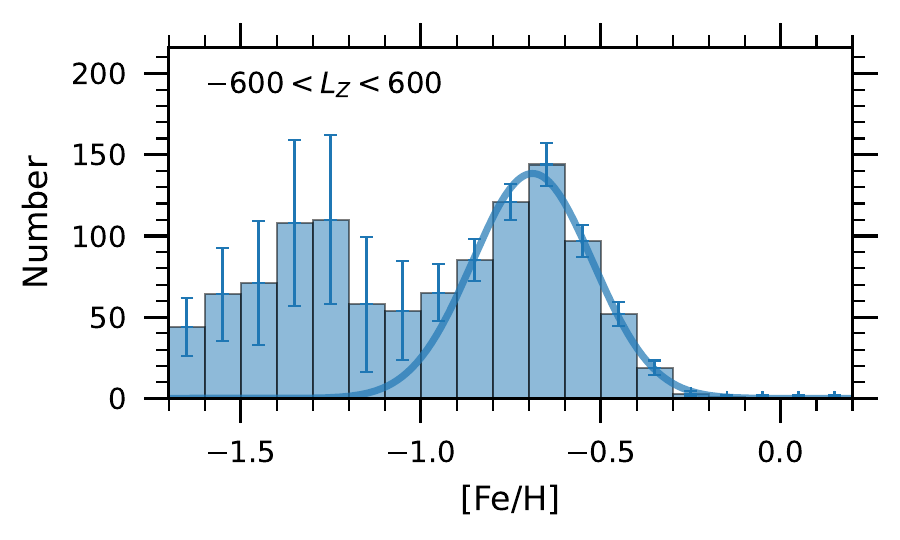}
    \caption{\textbf{SDSS}}
\end{subfigure}
\begin{subfigure}{8.4cm}
    \centering
    \includegraphics[width=\textwidth]{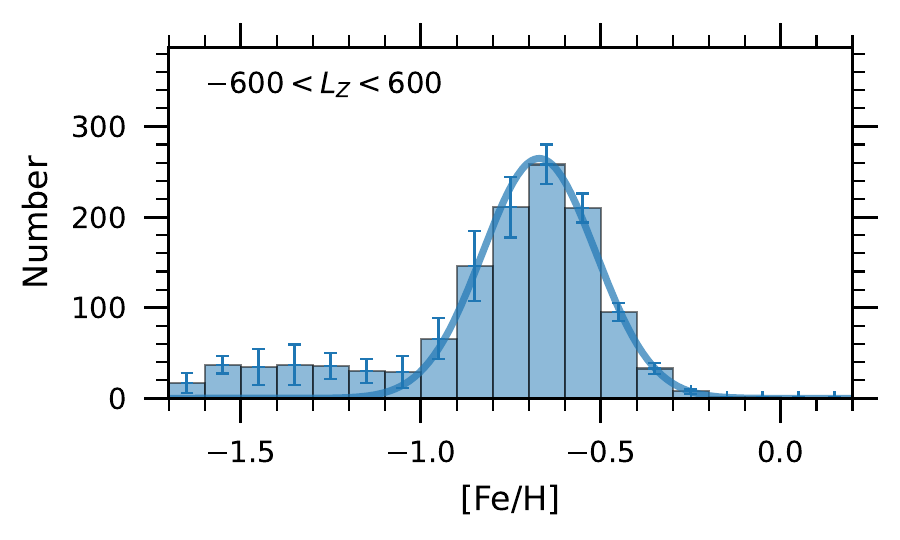}
    \caption{\textbf{LAMOST}}
\end{subfigure}
\caption{Metallicity distributions of high-$\alpha$ stars with $|L_Z| < 600~\kks$, representing a predominantly heated population. The [Fe/H] distribution was modelled with a single Gaussian (solid line). Error bars reflect uncertainties arising from variations in the $\alpha$-based division (see text).}
\label{fig:mdf_highafe}
\end{figure*}

\begin{figure*}
\centering
\begin{subfigure}{8.4cm}
    \centering
    \includegraphics[width=\textwidth]{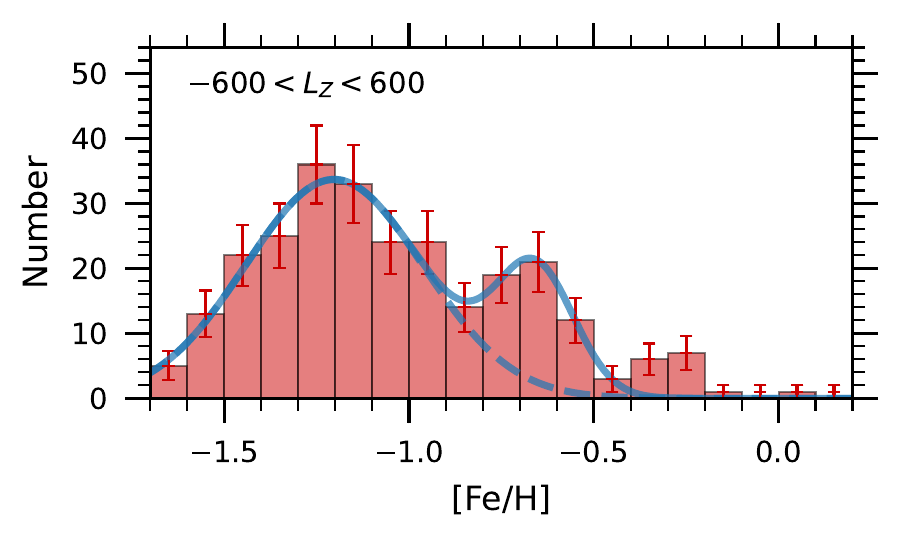}
    \caption{\textbf{GALAH}}
\end{subfigure}
\vspace{0.4cm}
\begin{subfigure}{8.4cm}
    \centering
    \includegraphics[width=\textwidth]{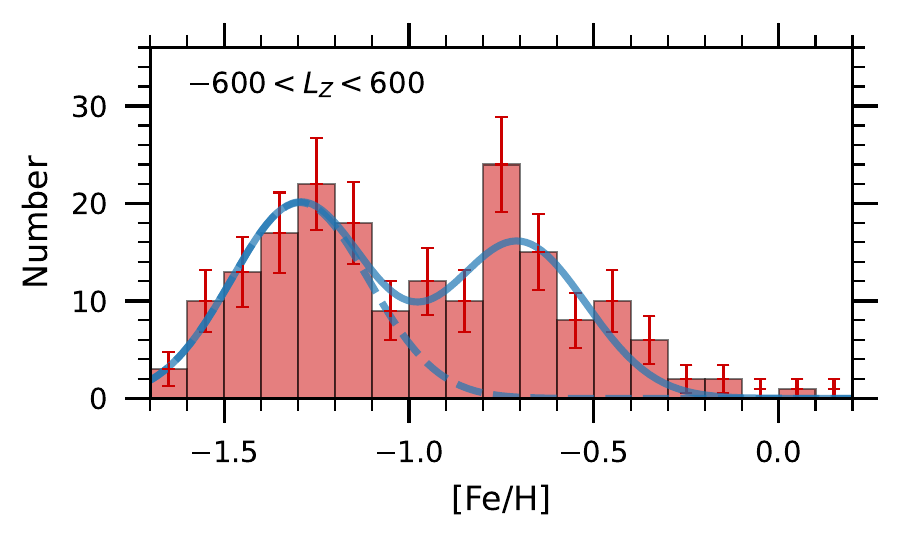}
    \caption{\textbf{APOGEE}}
\end{subfigure}
\begin{subfigure}{8.4cm}
    \centering
    \includegraphics[width=\textwidth]{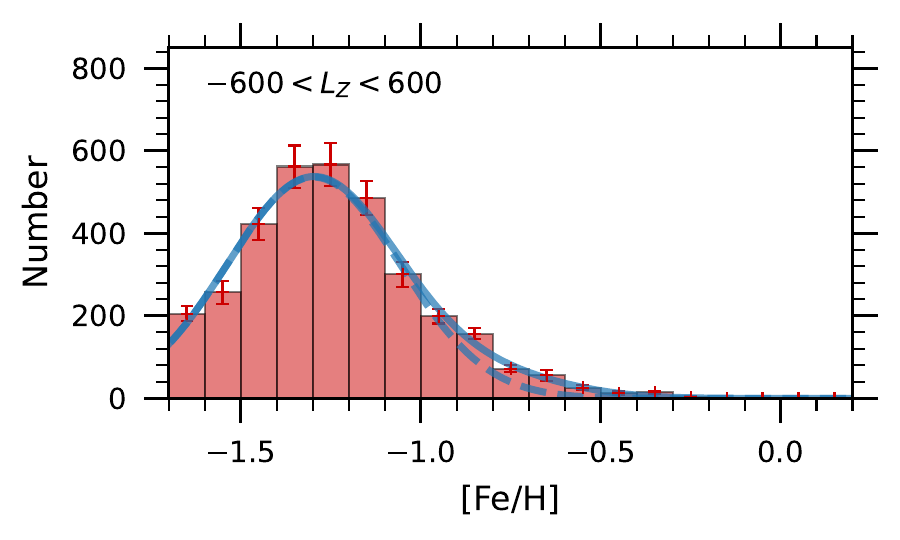}
    \caption{\textbf{SDSS}}
\end{subfigure}
\begin{subfigure}{8.4cm}
    \centering
    \includegraphics[width=\textwidth]{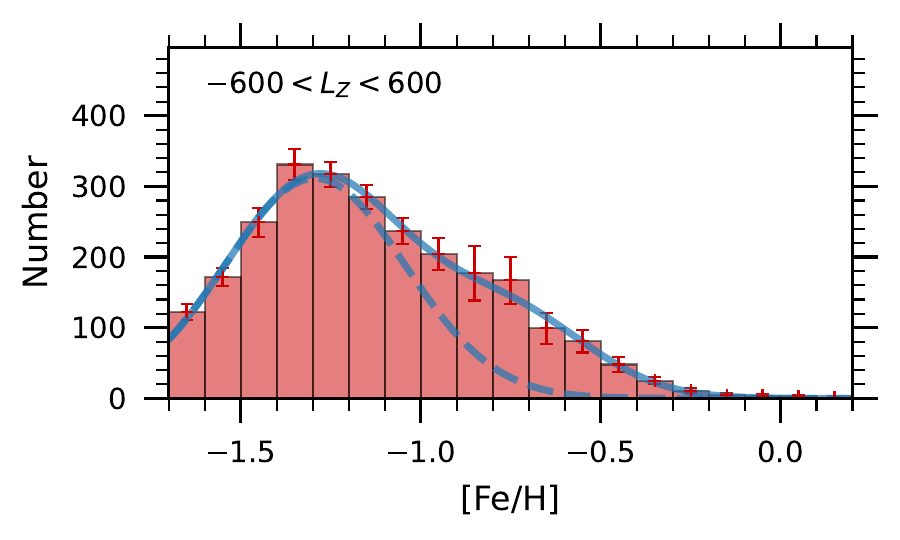}
    \caption{\textbf{LAMOST}}
\end{subfigure}
\caption{Metallicity distributions of low-$\alpha$ stars with $|L_Z| < 600~\kks$. Two Gaussian components were used to model the observed distribution: a broad one representing accreted stars from GSE (dashed line), and another capturing the secondary peak linked to the LAHN population. The sum of both components is shown as a solid line. Error bars reflect uncertainties due to variations in the $\alpha$-based classification (see text).}
\label{fig:mdf_lowafe}
\end{figure*}

Having established that the LAHN stars likely formed as a result of star formation triggered by the GSE merger, we estimated their relative fraction with respect to both the accreted GSE population and the heated-disc stars. This was achieved by decomposing the observed metallicity distributions from the spectroscopic samples for the high-$\alpha$ and low-$\alpha$ groups, respectively (see Fig.~\ref{fig:mdf}). We focused on the lower portion of the HPMS ($L_Z \sim 0\ \kks$), where the accreted, heated, and LAHN populations overlap, and adopted a relatively large bin size ($|L_Z| < 600\ \kks$) to ensure sufficient sample size for this analysis.

Figures~\ref{fig:mdf_highafe}--\ref{fig:mdf_lowafe} show the decomposition results for each of the four spectroscopic samples. Error bars indicate the uncertainties arising from the $\alpha$-based division (Appendix~\ref{sec:appendix1}). For the high-$\alpha$ stars (Fig.~\ref{fig:mdf_highafe}), which primarily belong to the heated-disc component, we modelled the metallicity distribution over the range $-1.0 < {\rm [Fe/H]} < -0.1$ with a single Gaussian function, shown as a solid line. While an extended metal-poor tail is evident, a significant portion of it may be due to uncertainties in the $\alpha$-based classification. The peak and width of the distributions are similar across the samples, with (mean, standard deviation) values of $(-0.60, 0.18)$, $(-0.59, 0.15)$, $(-0.69, 0.17)$, and $(-0.67, 0.16)$ from panel~(a) to (d), respectively. The small differences in [Fe/H] may reflect slight systematic variations in the metallicity calibration (see Sect.~\ref{sec:zp}).

For the low-$\alpha$ stars (Fig.~\ref{fig:mdf_lowafe}), we decomposed the [Fe/H] distribution into two Gaussian components over the range $-1.7 < {\rm [Fe/H]} < -0.4$, motivated by the presence of the LAHN stars identified in the high-resolution spectroscopic data sets (Sect.~\ref{sec:LAHN}). Even without this assumption, the GALAH and APOGEE samples clearly reveal a secondary peak at [Fe/H] $\approx-0.7$, associated with the LAHN stars, superimposed on a broader distribution centred at [Fe/H] $\approx-1.3$, corresponding to GSE. Reassuringly, the $\sim$1:2 population ratio between the primary (accreted) and secondary (LAHN) components over $-0.9 < {\rm [Fe/H]} < -0.4$ aligns with our chemical identification of the LAHN stars in the [Na/Fe]--[Fe/H] plane (Fig.~\ref{fig:abun}).

\begin{figure}
\centering
\includegraphics[width=7.8cm]{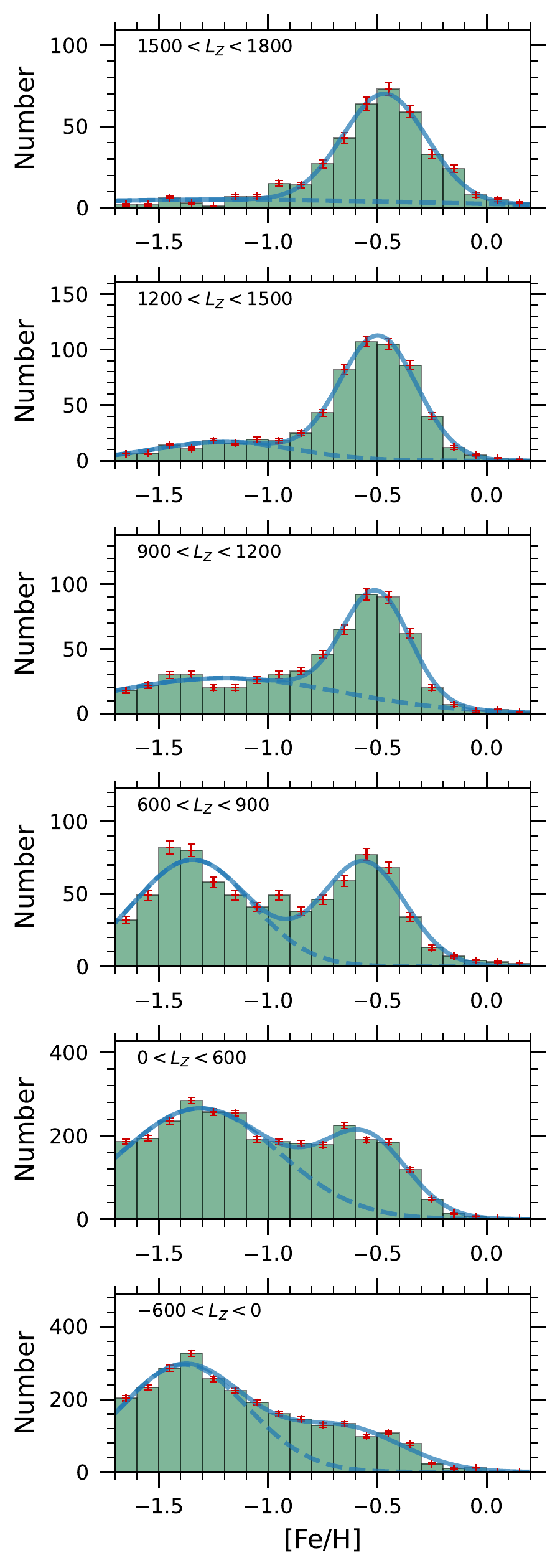}
\caption{Decomposition of the metallicity distribution from the photometric sample. The distribution was modelled using two Gaussian components: one peaking at [Fe/H]~$\sim -1.3$, representing the GSE debris, and a more metal-rich component reflecting a mixture of high-$\alpha$ (heated-disc) stars and the LAHN population. Error bars indicate uncertainties estimated via bootstrap resampling.}
\label{fig:mdf_phot}
\end{figure}

Following this approach, Fig.~\ref{fig:mdf_phot} presents the decomposition of metallicity distributions derived from photometric data, ranging from $L_Z = -600\ \kks$ to $1800\ \kks$. The error bars in the histograms were computed using bootstrap resampling. Unlike the spectroscopic samples, photometric data are limited by the absence of $\alpha$-element abundance measurements. However, since the LAHN population exhibits a [Fe/H] distribution nearly identical to that of the heated-disc stars, the overall distribution can still be effectively modelled with two Gaussian components: one representing accreted stars with a low-metallicity peak, and the other corresponding to the high-metallicity component that includes both heated-disc stars and the LAHN population. The fit to the photometric data using these two components is shown by the solid lines in Fig.~\ref{fig:mdf_phot}. The fitting was carried out independently within each $L_Z$ bin. Compared to the spectroscopic data, the best-fitting Gaussian components display broader distributions, primarily due to the larger [Fe/H] uncertainties in the photometric sample. At [Fe/H] $> -1$, the fraction of GSE stars decreases with increasing $L_Z$, while the combined fraction of the LAHN and heated-disc populations increases accordingly. This trend provides further support for the transition from accreted to in situ stellar populations along the HPMS.

\begin{figure}
\centering
\begin{subfigure}{8.7cm}
    \centering
    \includegraphics[width=\textwidth]{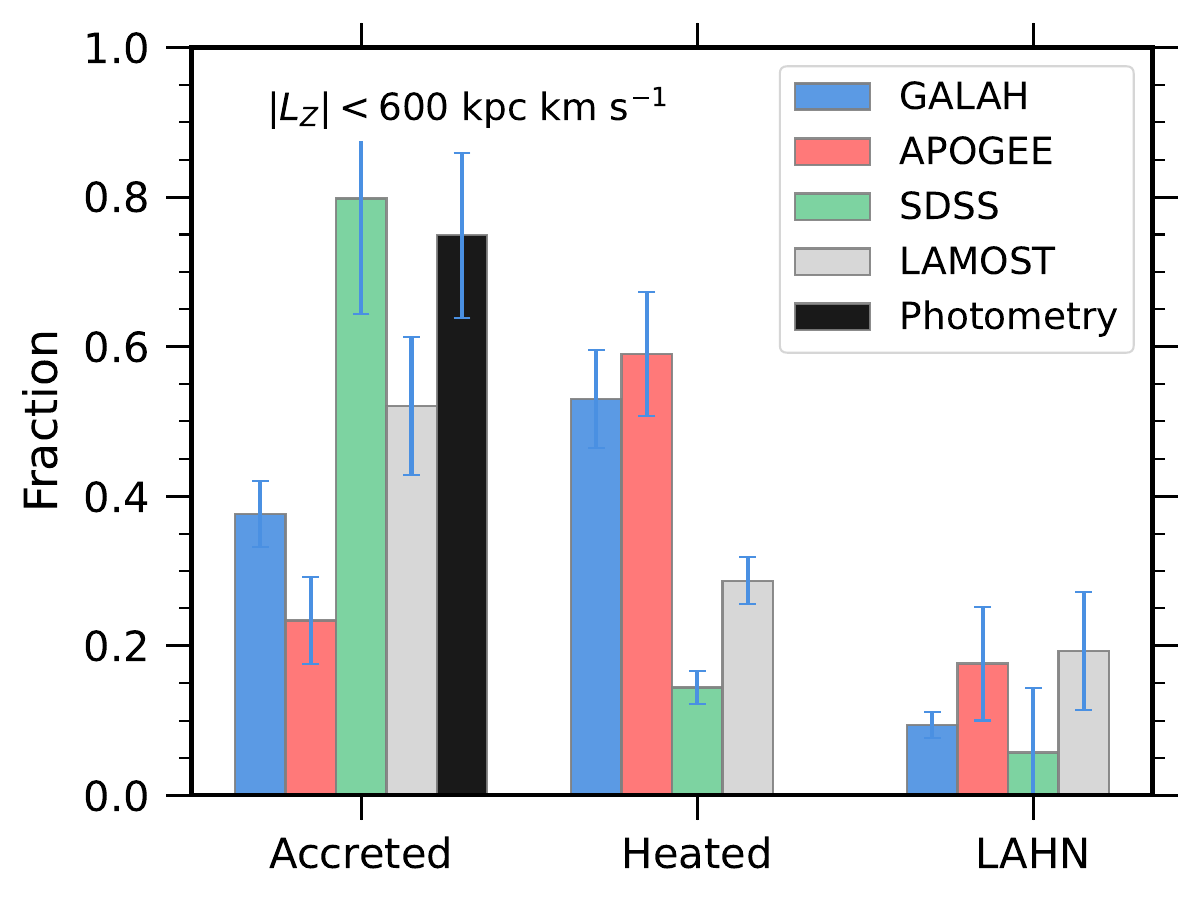}
    \caption{\textbf{Raw fractions}}
\vspace{0.6cm}
\end{subfigure}
\begin{subfigure}{8.7cm}
    \centering
    \includegraphics[width=\textwidth]{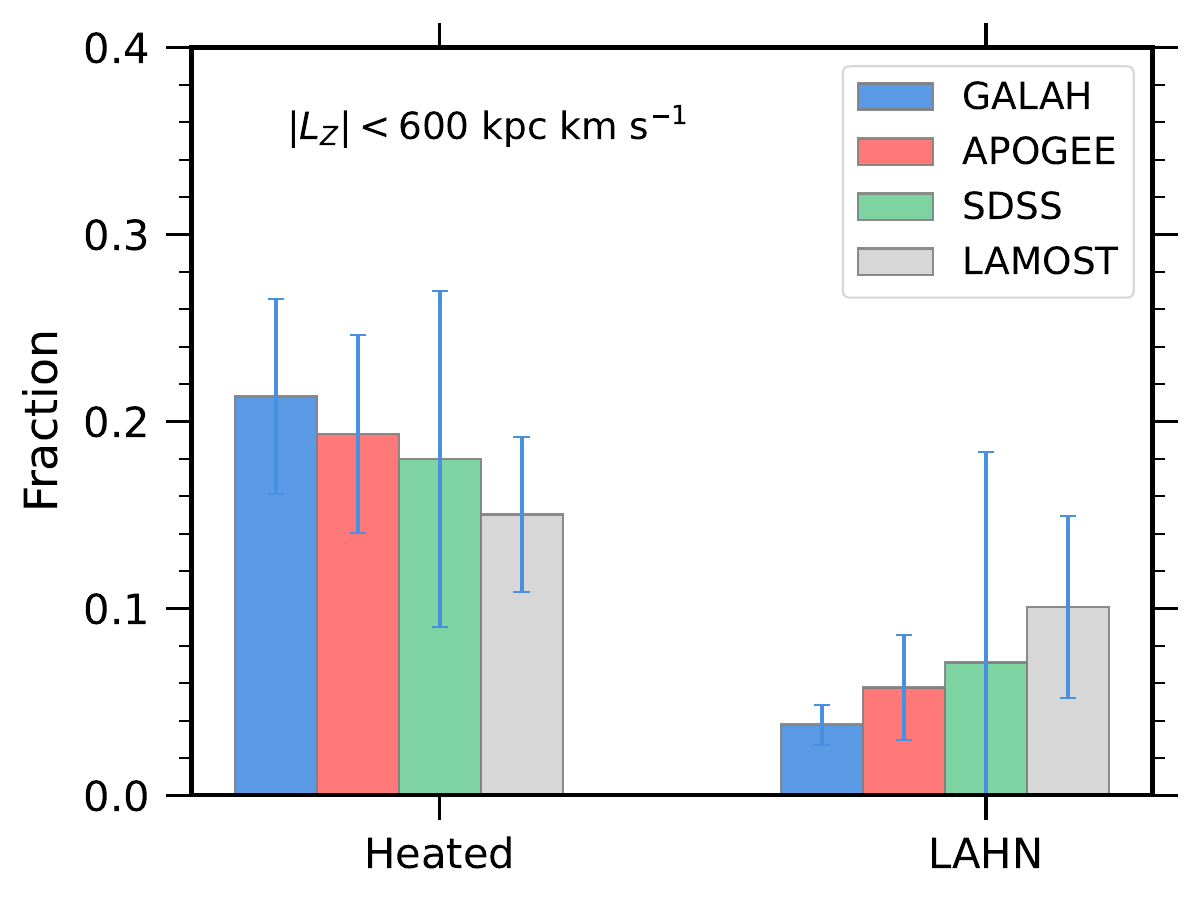}
    \caption{\textbf{Normalised fractions}}
\end{subfigure}
\caption{Fractions of individual stellar populations identified in the HPMS samples. Top: Raw fractions of the accreted, heated-disc, and LAHN populations, derived from Gaussian decomposition (Figs.~\ref{fig:mdf_highafe}--\ref{fig:mdf_phot}). For the photometric sample, only the accreted component is shown, because the heated and LAHN populations cannot be separated based on photometric metallicity estimates alone. Bottom: Normalised fractions of the heated and LAHN populations, scaled so that their combined contribution matches that derived from the photometric sample.}
\label{fig:frac}
\end{figure}

\begin{table}
\caption{Population fractions ($|L_Z| < 600\ \kks$)}
\label{tab:tab1}
\centering
\begin{tabular}{l c c c}
\hline\hline
 & Accreted & Heated & LAHN\tablefootmark{a} \\
Sample & ($\%$) & ($\%$) & ($\%$) \\
\hline
\multicolumn{4}{c}{Raw Fractions} \\
\hline
GALAH & $38\pm4$ & $53\pm6$ & $9\pm2$ \\
APOGEE & $23\pm6$ & $58\pm9$ & $19\pm7$ \\
SDSS & $79\pm17$ & $14\pm2$ & $7\pm9$ \\
LAMOST & $53\pm9$ & $29\pm3$ & $18\pm7$ \\
Photometry & $75\pm11$ & \multicolumn{2}{c}{$25\pm5$} \\
\hline
\multicolumn{4}{c}{Normalised Fractions} \\
\hline
GALAH & $75$ & $21\pm5$ & $4\pm1$ \\
APOGEE & $75$ & $19\pm5$ & $6\pm3$ \\
SDSS & $75$ &  $17\pm8$ & $8\pm11$ \\
LAMOST & $75$ & $15\pm4$ & $10\pm4$ \\
\hline
Weighted Mean & $75$ & $19\pm3$ & $6\pm3$ \\
\hline
\end{tabular}
\tablefoot{
\tablefoottext{a}{Low-$\alpha$, high-Na stars.}}
\end{table}

The estimated fractions of the accreted (GSE), heated-disc, and LAHN populations in the HPMS samples are provided in the upper part of Table~\ref{tab:tab1} (labelled `raw fractions') and displayed in the top panel of Fig.~\ref{fig:frac}. The uncertainties in these fractions were derived by propagating the fitting errors and accounting for covariances in the decomposition process. Notably, the raw fraction of the LAHN population remains relatively constant, ranging from 7\% to 19\% across all samples. For the LAMOST sample, we took advantage of the larger dataset to subdivide it into two $L_Z$ bins (see Appendix~\ref{sec:appendix3}); however, no clear trend was detected in the population fractions across these bins.

The estimated contribution of the LAHN stars relative to the accreted component is significantly lower than that reported by \citet{lee:23}, who analysed the same SDSS and LAMOST datasets used in this study. They identified roughly equal proportions of the starburst (i.e.\ LAHN-like) and GSE populations among low-$\alpha$ stars with $e > 0.7$, but found only a negligible contribution of starburst stars among high-$\alpha$ stars, which were dominated by the heated-disc population. However, the population divisions in \citeauthor{lee:23} were primarily phenomenological in nature, based on a statistical decomposition of observed trends in orbital inclination and radial velocity dispersion with metallicity. Moreover, their analysis did not incorporate light-element abundances, and thus the stars they classified as starburst-origin may not correspond to the chemically defined LAHN population in this work. Consequently, the assignment of stars to specific populations was inherently model-dependent, leaving room for re-interpretation of the resulting population fractions.

As is shown in Table~\ref{tab:tab1}, the different spectroscopic samples yield significantly varying population fractions, largely due to sample selection biases. For instance, the SDSS sample shows the highest contribution from the GSE population, whereas the APOGEE sample shows the lowest. This trend is reversed for the heated-disc population. To correct for the sampling bias inherent in each spectroscopic sample, we adopted the total fraction of the heated and LAHN populations inferred from the photometric sample ($25\%\pm5\%$) as a reference for all spectroscopic datasets. The resulting bias-corrected values are presented in the lower part of Table~\ref{tab:tab1} (labelled `normalised fractions') and illustrated in the bottom panel of Fig.~\ref{fig:frac}. Following this correction, the relative contributions of the heated and LAHN populations show good agreement across the various spectroscopic datasets. The resulting weighted mean fractions for stars within $|L_Z| < 600\ \kks$ and $\zmax > 2$~kpc are $19\%\pm3\%$ for the heated population and $6\%\pm3\%$ for the LAHN population in our HPMS samples.

Photometric metallicity estimates are also subject to sample biases. In the most restricted case, with volume- and mass-limited cuts \citep[e.g.][]{an:13}, there is a known bias against metal-rich main-sequence stars, which are intrinsically fainter than metal-poor stars of the same colour and are thus underrepresented near the survey’s brightness limit. However, the sample shown in Fig.~\ref{fig:mdf_phot} is less susceptible to such biases owing to our selection of nearby, relatively bright stars with reliable astrometric measurements. Furthermore, the ratio between the heated and LAHN populations ($\sim$3:1) is expected to remain robust, as both populations exhibit similar metallicities.

\section{Summary and discussion}\label{sec:discussion}

In this study, we investigated the origin and characteristics of the HPMS \citep{paper4}, identifying a distinct population of LAHN stars that likely formed during the GSE merger event. Using high-resolution spectroscopic data, we demonstrated that this population exhibits unique chemical signatures -- particularly enhanced Na and Al abundances -- distinguishing them from both the accreted stars of GSE and the heated-disc population. Their spatial and orbital properties further suggest that they formed from gas clouds originating in the GSE progenitor galaxy, which underwent rapid star formation before being mixed with clouds in the MW. Our findings are supported by numerical simulations of a MW-like galaxy, which reveal a burst of in situ star formation triggered by the merger. The simulated starburst populations exhibit spatial, dynamical, and chemical properties consistent with our observational results, reinforcing the hypothesis that the LAHN stars are direct descendants of such an event. Through decomposition of the metallicity distribution in the HPMS samples, we estimated that the LAHN population constitutes approximately 6\% of the local high proper-motion stars with GSE-like kinematics, with the remaining fractions attributed to the GSE debris ($\sim75\%$) and dynamically heated-disc stars ($\sim19\%$). These results provide strong evidence that the gas-rich merger not only deposited ex situ stars into the MW, but also triggered a significant episode of in situ star formation, playing a crucial role in shaping the early MW.

\subsection{Hypothesis-driven perspective on Eos}

We demonstrated that the HPMS, traced by high proper-motion stars, includes a population of stars formed during an episode of intense star formation. However, not all stars within the HPMS are direct products of this process. Rather, the HPMS traces a broader sequence of stars that were accreted, dynamically modified, or formed during the interaction between the GSE and the MW. It highlights the intricate interplay of accretion, heating, and star formation triggered by the GSE merger, offering deeper insights into how major Galactic collisions influenced the chemical and kinematic properties of stellar populations.

In this sense, our study offers a complementary perspective on the Eos group identified by \citet{myeong:22}. Using Gaussian mixture models with elemental abundances from APOGEE DR17 and GALAH DR3, combined with orbital energy, they found that Eos exhibits higher [Al/Fe] than GSE stars and is more tightly bound to the MW. Notably, they proposed that Eos bridges the GSE population at low metallicity and the low-$\alpha$ thin disc at higher metallicity. While the LAHN stars share many chemical similarities with Eos, our study adopted an independent approach, identifying a distinct stellar population likely linked to merger-driven star formation. The key differences relative to Eos can be summarised as follows:

\begin{enumerate}[label=\roman*)]

\item The primary distinction lies in the methodological approach adopted. \citet{myeong:22} employed an unsupervised machine learning method to identify substructure without invoking a specific formation scenario. In contrast, our study was hypothesis-driven: we searched for stars that could plausibly have formed through merger-driven star formation by focusing on the HPMS and identifying those with chemical and dynamical properties expected for a population formed during the GSE merger event.

\item Within our hypothesis-driven framework, we identified a broader eccentricity distribution among LAHN stars, spanning $e \sim 0.55$ to $1.0$, with one possible outlier at $e \sim 0.35$. In contrast, \citet{myeong:22} applied a selection cut of $e > 0.85$, effectively restricting their sample to highly radial orbits. Our results imply that stars with similar chemical characteristics to Eos are not limited to extreme orbits, and that imposing strict eccentricity thresholds may overlook a substantial fraction of this chemically distinct population.

\item Our analysis revealed that the LAHN population occupies more spatially confined orbits than GSE, with smaller apogalacticons and vertical excursions. While Eos is also known to exhibit tightly bound orbits \citep{matsuno:24}, our results revealed an additional layer of dynamical structure: a systematic correlation between orbital eccentricity and $\zmax$ (see below).

\item Our study uncovered a clear bimodality in the $L_Z$ distribution of low-$\alpha$, high-Na stars. One group, referred to as the LAHN stars, is concentrated around $L_Z \sim 0\ \kks$, while the other, with LAHN-like chemical abundances, overlaps with the high-$L_Z$ regime typical of disc stars. The notable absence of such stars between these two $L_Z$ groups suggests discrete episodes of star formation, likely reflecting a non-continuous inflow of star-forming gas.

\end{enumerate}

\subsection{The precursor to the merger-driven starburst}

Our analysis of the LAHN population provided new insights into the aftermath of the GSE merger and its role in the early evolution of the MW. Although modest in number, these stars exhibit notable chemical homogeneity and intermediate abundance patterns, offering critical clues about the origin, timescale, and environment of their formation.

For the LAHN stars in the GALAH sample, the $1\sigma$ dispersion in [Fe/H] is 0.1~dex, while [Na/Fe] and [Al/Fe] show dispersions as small as $\sim$0.1~dex and 0.2~dex, respectively (Figs.~\ref{fig:abun} and \ref{fig:mdf_lowafe}). The dispersion in [Fe/H] is about half that observed in the heated-disc component. Such tight chemical abundance patterns are difficult to achieve in systems with extended, multi-phase star formation, where progressive chemical enrichment or incomplete mixing of the interstellar medium could produce significant abundance spreads \citep[e.g.][]{escala:18}. Star formation in a turbulent disc environment is also plausible, but in that case, these stars would be expected to have more circularised orbits \citep[e.g.][]{brook:04,sales:09} rather than GSE-like kinematics with extreme eccentricities.

Instead, the observed chemical homogeneity points towards a scenario in which rapid star formation occurred from a relatively compact, well-mixed gas reservoir. This may align with the fact that the LAHN stars comprise only a modest fraction of the MW's stellar populations. Cosmological zoom-in simulations show that stars can form in short, intense bursts during merger events \citep[e.g.][]{renaud:14,moreno:15,ma:17,sparre:17,yu:21,hirai:22,renaud:22,hirai:24}. These bursts typically last less than a few hundred million years. Based on our estimated population fraction -- roughly 1 LAHN star for every 10 GSE stars -- we inferred a total stellar mass of $\sim5 \times 10^7\ M_\odot$ for the LAHN population, assuming a stellar mass of $5 \times 10^8\ M_\odot$ for the GSE progenitor \citep{naidu:21}. This implies a star-formation rate of the order of $0.1\ M_\odot\ \mathrm{yr}^{-1}$ for the gas clouds that gave rise to the LAHN stars.

The star formation rate associated with the LAHN population appears modest, likely comparable to that of the Central Molecular Zone in the present-day MW \citep{an:11}, which harbours a dense gas reservoir of $\sim5 \times 10^7\ M_\odot$ \citep{pierceprice:00}. Still, the inferred star-formation rate remains well below the threshold typically associated with classical starbursts \citep[$\sim10$--$100\ M_\odot\ \mathrm{yr}^{-1}$;][]{kennicutt:98}. It is also significantly lower than the peak rates of MW analogues at cosmic noon, estimated to be on the order of a few tens of $M_\odot\ \mathrm{yr}^{-1}$ \citep{papovich:15}. These considerations suggest that, although the star formation responsible for this population was locally significant, it did not dominate the global star-forming activity in the early MW.

Indeed, compact, high-intensity star-forming regions have also been observed in high-redshift galaxies. For example, \citet{liu:24} reported a dusty star-forming clump in a $z \sim 1.5$ galaxy using JWST and ALMA, with properties consistent with a short-duration, high-efficiency burst. Likewise, the gravitationally lensed images of a galaxy at $z \sim 1.4$ \citep{mowla:22,claeyssens:23} reveal compact clumps within its main galaxy that may represent either young globular clusters or sites of localised starbursts. The chemically coherent LAHN population we identified may represent a fossil record of such bursty, clump-scale star formation common at high redshift.

In our localised, burst-driven formation scenario, the LAHN stars formed from metal-enriched gas directly accreted from the GSE progenitor. This interpretation is supported by their mean metallicities, which lie at the upper end of the metallicity distribution observed in GSE stars. In addition, the increased gas density, or shocks induced by the merger, likely triggered a brief yet intense episode of star formation within the cloud, enhancing $\alpha$- and light-element (Na and Al) abundances through core-collapse SNe. However, given the modest overall star-formation rate, the total yields of these elements would have been lower than those produced in the high-$\alpha$ disc population. This naturally explains why the elemental abundances of the LAHN stars occupy an intermediate regime between those of the GSE stars and the high-$\alpha$, heated-disc population. A more intricate scenario involving pristine gas inflows from the intergalactic medium or the outer disc remains plausible \citep[e.g.][]{renaud:21}, implying a distinct enrichment history and potentially diverse origins. Nevertheless, the similar kinematics of these stars to the GSE debris argue against a purely outer disc or intergalactic origin.

In addition, the timing of star formation can be inferred from the orbital properties of the LAHN stars. As is shown in Fig.~\ref{fig:asp}, most of them exhibit a tight correlation between orbital eccentricity and $\zmax$, such that more eccentric orbits tend to reach higher vertical distances. As is further demonstrated in Appendix~\ref{sec:appendix2}, high-$\alpha$ stars also follow a similar relation, which likely arises from stronger vertical perturbations during mergers. The fact that the LAHN stars align with this trend suggests that they formed during the same episode of dynamical disturbance, when infalling gas clouds from the dwarf galaxy collided with the primordial disc of the MW.

In the context of a merger origin for the LAHN population, Omega Centauri ($\omega$~Cen) serves as a valuable analog. $\omega$ Cen exhibits a large internal spread in metallicity, extending up to [Fe/H] $\sim-0.7$, and is widely regarded as the remnant nucleus of a disrupted dwarf galaxy \citep{lee:99,pancino:00,bekki:03}. These characteristics suggest a prolonged enrichment history, consistent with multiple episodes of star formation and self-enrichment within a deep gravitational potential well. Notably, the most metal-rich stars in $\omega$~Cen, with [Fe/H] $\sim-0.6$ \citep{johnson:10,an:17}, exhibit a wide range of $\alpha$-element abundances, raising the possibility that some of our LAHN stars could trace the nuclear component of a progenitor dwarf galaxy.

Indeed, $\omega$~Cen members display elevated [Na/Fe] values ($\sim0.9$~dex) in the highest-metallicity group (${\rm [Fe/H]} > -0.9$), compared to the majority of the more metal-poor stars in the cluster \citep{johnson:10}. However, a key distinction lies in the detailed trends of the abundance ratios. While our sample of low-$\alpha$ stars includes a group with distinctly elevated [Na/Fe] (i.e.\ the LAHN group) atop a decreasing [Na/Fe] -- [Fe/H] trend seen in GSE stars (Fig.~\ref{fig:abun}), the stars in $\omega$~Cen follow a monotonically increasing [Na/Fe] trend with metallicity (see their Fig.~10). The [Al/Fe] abundances in $\omega$~Cen also display a more complex pattern: the metal-poor stars split into two distinct branches, while the most metal-rich stars show intermediate [Al/Fe] values, features not observed in our low-$\alpha$ sample (Figs.~\ref{fig:abun} and \ref{fig:abun2}). These contrasting chemical trends suggest that, despite superficial similarities, the LAHN population and $\omega$~Cen followed distinct chemical enrichment pathways. This likely reflects the effect of enhanced star-forming activity in the aftermath of the GSE merger, which led to greater contributions from core-collapse SN(e) and the resulting chemical enrichment.

Following the intense but short-lived formation episode of the LAHN stars, star formation may have temporarily ceased, as suggested by the absence of low-$\alpha$ stars at intermediate $L_Z$ (Fig.~\ref{fig:galah_lz}). The residual gas left behind after the merger may have funnelled into the inner MW, where it mixed with pre-existing clouds in the primordial disc and eventually fuelled the formation of subsequent stellar generations \citep[e.g.][]{buck:20}. A possible signature of this process is found in stars along the HPMS that exhibit similar chemical abundances to LAHN stars but have larger $L_Z$ values (grey crosses in Fig.~\ref{fig:asp}). These stars likely formed from gas clouds that acquired significant angular momentum during the merger process, possibly through tidal torques \citep[e.g.][]{barnes:02, brook:12}.

Within this framework, the starburst episode identified by \citet{ciuca:24} may represent a later stage in a merger-driven, multi-phase star-formation history in the early MW. While their starburst sample spans a metallicity range comparable to that of our LAHN stars, it occupies a distinct kinematic regime, with stars predominantly at $L_Z > 1000\ \kks$, significantly larger than those of the LAHN population. This distinction suggests that the discontinuity in metallicity and [$\alpha$/Fe] trends with stellar age reported by \citeauthor{ciuca:24} likely reflects a starburst event that occurred after the merger-driven gas clouds had settled into the pre-existing disc. In contrast, the LAHN stars may mark an earlier, dynamically chaotic phase of star formation, initiated by the same merger event. Taken together, these merger-driven populations may signify the transition from bursty to steady star formation in the MW, setting the stage for the subsequent, more extended phases of inside-out disc growth \citep[e.g.][]{ma:17,yu:21,hirai:22,mccluskey:24}.

\section{Conclusions}

In this study, we analysed a set of high proper-motion stars from photometric and spectroscopic surveys, selected to trace the aftermath of the GSE merger. Within these datasets, we identified and characterised a chemically and kinematically distinct stellar population, which we interpreted as the product of moderately intense star formation triggered by the merger. Our analysis leads to the following conclusions:

\begin{itemize}

\item A subset of metal-rich ([Fe/H] $\approx-0.6$) stars with GSE-like kinematics exhibit low-$\alpha$ and high-Na abundances (LAHN stars). Their chemically distinct abundance patterns distinguish them from both the accreted GSE debris and the heated-disc population (Splash).

\item While their chemical properties resemble those of the Eos population, the LAHN stars span a significantly broader range of orbital eccentricities ($0.5 \la e \la 1$), suggesting that Eos may represent the high-eccentricity tail of a more extensive population.

\item The similarity in orbital properties ($L_Z$ and $e$) between the LAHN stars and the GSE debris suggests that the LAHN stars may have formed from gas accreted from the GSE progenitor dwarf galaxy. Their present-day spatial-orbital structure, specifically the coupling between $\zmax$ and $e$, also implies that they may have been subsequently displaced into the halo by dynamical heating.

\item Nonetheless, the LAHN stars are more concentrated in the inner MW than the GSE stars. This finding is consistent with theoretical predictions that gas-rich mergers at high redshift can lead to the formation of stars in the inner regions of a host galaxy. Their elevated Na and/or Al abundances additionally support an in situ origin, indicating that these stars likely formed during active star formation rather than being directly accreted from a dwarf galaxy.

\item The homogeneous chemical abundances of LAHN stars, together with their modest population ratio ($\sim1:10$) relative to the accreted GSE stars, suggest that they formed within spatially compact gas clouds, analogous to clumpy star-forming regions observed in high-redshift galaxies.

\item The modest population fraction also suggests that the star formation immediately following the GSE merger did not play a dominant role in the early assembly of the MW. Instead, the LAHN population may represent an early and localised episode of star formation, providing observational insight into the conditions that triggered starburst activity during cosmic noon.

\end{itemize}

Although the contribution of gas accreted from the GSE progenitor to the formation of the low-$\alpha$ disc is not yet fully constrained, our findings point to a scenario in which merger-driven star formation influenced the early assembly of the thin disc. Future work combining precise stellar ages with observed chemical and dynamical properties of stars will further clarify the connection between this merger-driven star formation and the assembly of the Galactic disc system. In particular, upcoming surveys such as the Legacy Survey of Space and Time \citep[LSST;][]{ivezic:19} will enable the detection of ancient merger-driven starburst populations across a wider spatial and dynamical range, thereby offering new insight into their role in the evolution of the MW.

\begin{acknowledgements}

D.A.\ acknowledges support provided by the National Research Foundation (NRF) of Korea grant funded by the Ministry of Science and ICT (No.\ 2021R1A2C1004117). Y.S.L.\ acknowledges support from the NRF of Korea grant (RS-2024-00333766). Y.H.\ acknowledges support from the JSPS KAKENHI Grant Numbers JP22KJ0157, JP25H00664, and	JP25K01046. Numerical computations and analysis were carried out on Cray XC50 at the Center for Computational Astrophysics, National Astronomical Observatory of Japan. T.C.B.\ acknowledges partial support for this work from grant PHY 14-30152; Physics Frontier Center/JINA Center for the Evolution of the Elements (JINA-CEE), and OISE-1927130: The International Research Network for Nuclear Astrophysics (IReNA), awarded by the US National Science Foundation.\\

Funding for the Sloan Digital Sky Survey IV has been provided by the Alfred P. Sloan Foundation, the U.S. Department of Energy Office of Science, and the Participating Institutions. SDSS-IV acknowledges support and resources from the Center for High Performance Computing at the University of Utah. The SDSS website is www.sdss4.org. SDSS-IV is managed by the Astrophysical Research Consortium for the Participating Institutions of the SDSS Collaboration including the Brazilian Participation Group, the Carnegie Institution for Science, Carnegie Mellon University, Center for Astrophysics | Harvard \& Smithsonian, the Chilean Participation Group, the French Participation Group, Instituto de Astrof\'isica de Canarias, The Johns Hopkins University, Kavli Institute for the Physics and Mathematics of the Universe (IPMU) / University of Tokyo, the Korean Participation Group, Lawrence Berkeley National Laboratory, Leibniz Institut f\"ur Astrophysik Potsdam (AIP), Max-Planck-Institut f\"ur Astronomie (MPIA Heidelberg), Max-Planck-Institut f\"ur Astrophysik (MPA Garching), Max-Planck-Institut f\"ur Extraterrestrische Physik (MPE), National Astronomical Observatories of China, New Mexico State University, New York University, University of Notre Dame, Observat\'orio Nacional / MCTI, The Ohio State University, Pennsylvania State University, Shanghai Astronomical Observatory, United Kingdom Participation Group, Universidad Nacional Aut\'onoma de M\'exico, University of Arizona, University of Colorado Boulder, University of Oxford, University of Portsmouth, University of Utah, University of Virginia, University of Washington, University of Wisconsin, Vanderbilt University, and Yale University.\\

This work made use of the Fourth Data Release of the GALAH Survey \citep{buder:25}. The GALAH Survey is based on data acquired through the Australian Astronomical Observatory, under programmes: A/2013B/13 (The GALAH pilot survey); A/2014A/25, A/2015A/19, A2017A/18 (The GALAH survey phase 1); A2018A/18 (Open clusters with HERMES); A2019A/1 (Hierarchical star formation in Ori OB1); A2019A/15, A/2020B/23, R/2022B/5, R/2023A/4, R2023B/5 (The GALAH survey phase 2); A/2015B/19, A/2016A/22, A/2016B/10, A/2017B/16, A/2018B/15 (The HERMES-TESS programme); A/2015A/3, A/2015B/1, A/2015B/19, A/2016A/22, A/2016B/12, A/2017A/14, A/2020B/14 (The HERMES K2-follow-up programme); R/2022B/02 and A/2023A/09 (Combining asteroseismology and spectroscopy in K2); A/2023A/8 (Resolving the chemical fingerprints of Milky Way mergers); and A/2023B/4 (s-process variations in southern globular clusters). We acknowledge the traditional owners of the land on which the AAT stands, the Gamilaraay people, and pay our respects to elders past and present. This paper includes data that has been provided by AAO Data Central (datacentral.org.au).\\

Guoshoujing Telescope (the Large Sky Area Multi-Object Fiber Spectroscopic Telescope LAMOST) is a National Major Scientific Project built by the Chinese Academy of Sciences. Funding for the project has been provided by the National Development and Reform Commission. LAMOST is operated and managed by the National Astronomical Observatories, Chinese Academy of Sciences.

\end{acknowledgements}

\begin{appendix}
\onecolumn

\section{Empirical division of $\alpha$ sequences}\label{sec:appendix1}

\begin{figure*}[h]
\centering
\begin{subfigure}{16cm}
\centering
\includegraphics[width=\textwidth]{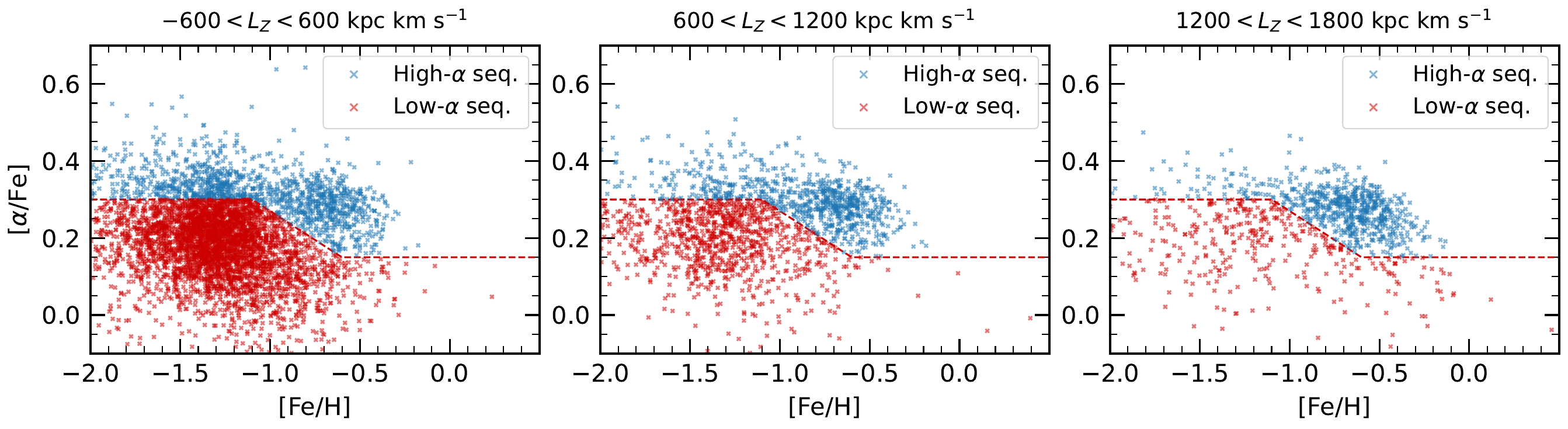}
\caption{\textbf{SDSS}}
\end{subfigure}
\begin{subfigure}{16cm}
\centering
\includegraphics[width=\textwidth]{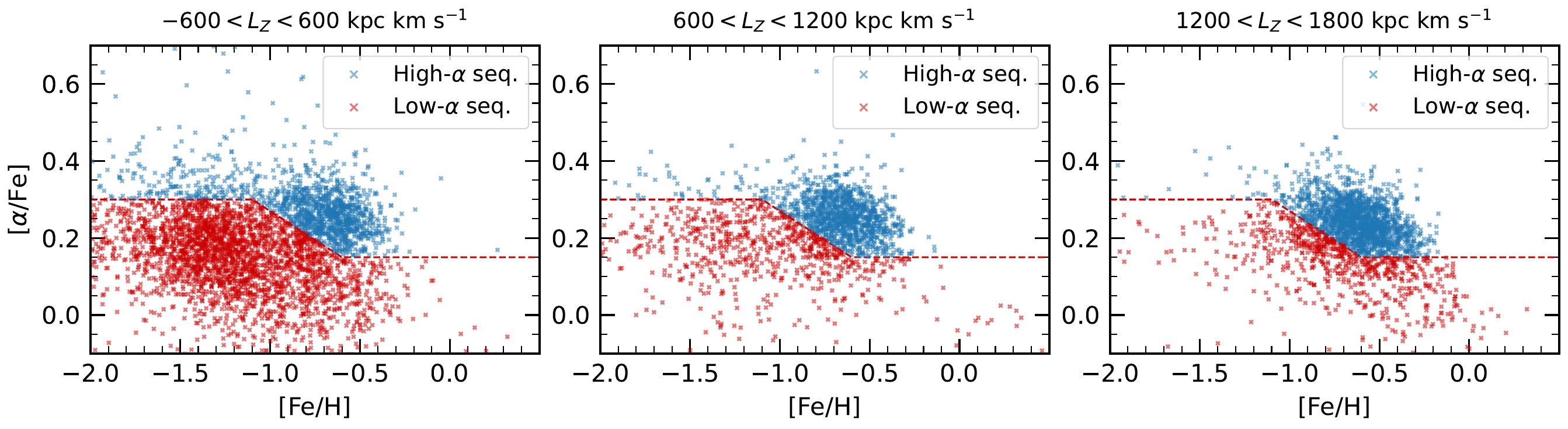}
\caption{\textbf{LAMOST}}
\end{subfigure}
\begin{subfigure}{16cm}
\centering
\includegraphics[width=\textwidth]{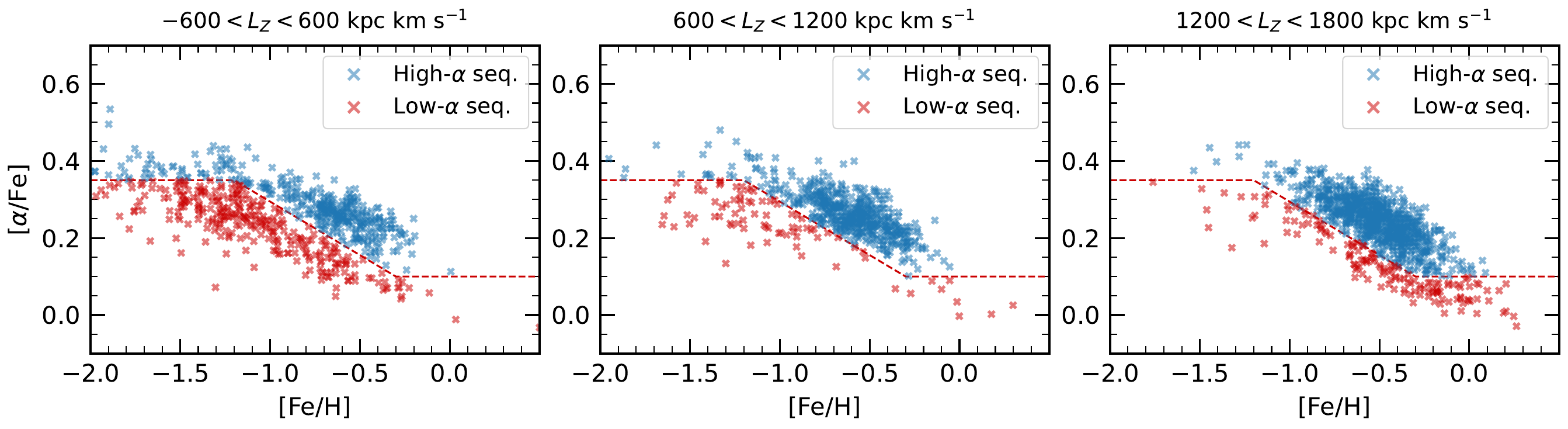}
\caption{\textbf{GALAH}}
\end{subfigure}
\begin{subfigure}{16cm}
\centering
\includegraphics[width=\textwidth]{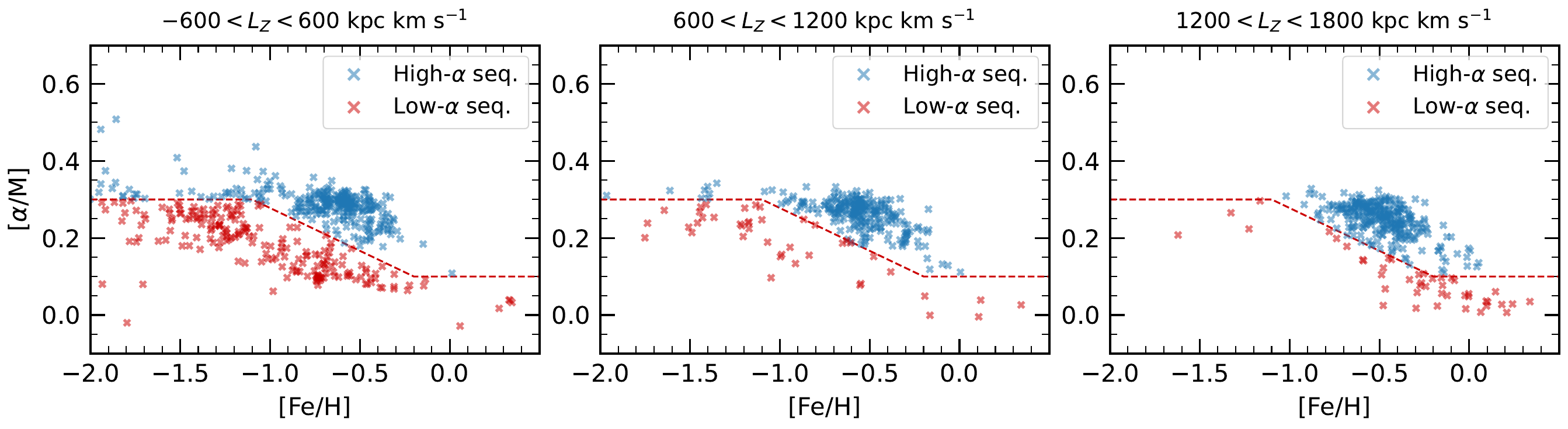}
\caption{\textbf{APOGEE}}
\end{subfigure}
\caption{Distribution of [$\alpha$/Fe] versus [Fe/H] for the spectroscopic samples used in this study. The dashed red lines, derived from these datasets by examining systematic variations across three $L_Z$ bins, were used to separate the high-$\alpha$ and low-$\alpha$ sequences within each sample.}
\label{fig:afe}
\end{figure*}

Figure~\ref{fig:afe} illustrates the separation between the low- and high-$\alpha$ sequences for the spectroscopic samples used in this study. The [Fe/H] vs. [$\alpha$/Fe] distributions are shown across three $L_Z$ bins, selected to highlight different stellar populations and their varying contributions. The high-$\alpha$ sequence spans the entire $L_Z$ range, corresponding to thick-disc stars at high $L_Z$ and heated-disc stars at low $L_Z$. In contrast, the low-$\alpha$ sequence consists mainly of accreted GSE stars at low $L_Z$ and thin-disc stars at high $L_Z$. A sparsely populated region at intermediate $L_Z$ provided an empirical basis for defining the division between the two sequences.

\FloatBarrier

\section{Spatial and orbital properties of high-$\alpha$ stars}\label{sec:appendix2}

\begin{figure*}[h]
\centering
\includegraphics[width=14cm]{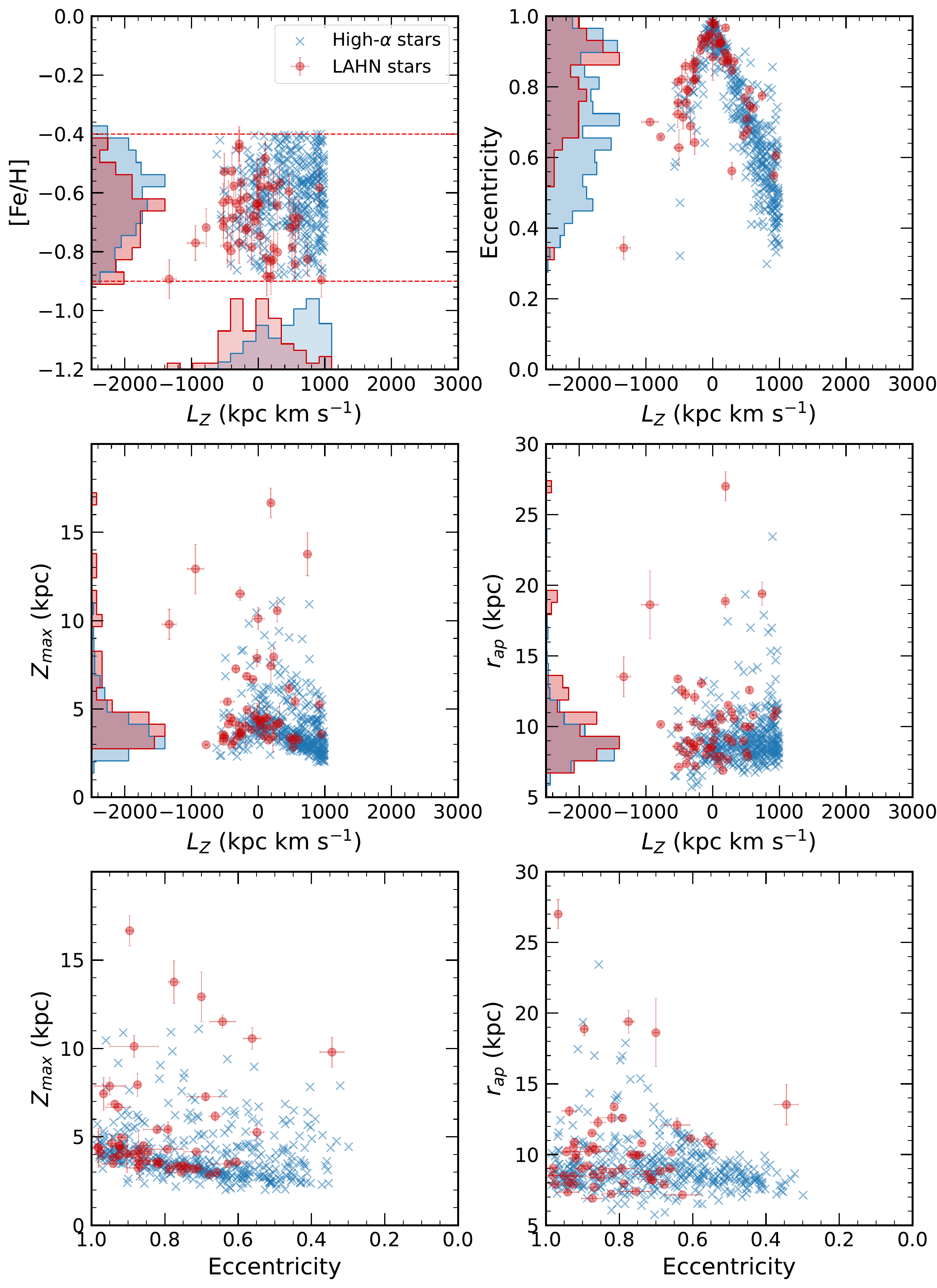}
\caption{Same as Fig.~\ref{fig:asp}, but comparing spatial and orbital properties of  high-$\alpha$ stars (blue crosses) and the LAHN population (red circles).}
\label{fig:asp2}
\end{figure*}

Figure~\ref{fig:asp2} presents the spatial and orbital properties of the high-$\alpha$ population (blue crosses), which were not included in Fig.~\ref{fig:asp}. For reference, the same set of LAHN stars from Fig.~\ref{fig:asp} is shown as red circles. The same selection criteria were applied, including $\zmax > 2$~kpc and $-0.9 < {\rm [Fe/H]} < -0.4$. Additionally, only stars with $L_Z < 1000\ \kks$ were included, where the high-$\alpha$ and LAHN populations overlap. As is shown in the bottom two panels, both populations exhibit a tight correlation between orbital eccentricity and $\zmax$, while the correlation between eccentricity and $r_{\rm ap}$ is less pronounced.

\FloatBarrier

\section{Sub-division of the LAMOST sample}\label{sec:appendix3}

Given the sufficient number of stars in the LAMOST sample, we subdivided the data into two $L_Z$ bins for the decomposition of its metallicity distribution: $-600 < L_Z < 0\ \kks$ and $0 < L_Z < 600\ \kks$. For the high-$\alpha$ stars, we applied a single Gaussian fit, as in the main sample (Fig.~\ref{fig:mdf_highafe}). No significant variation was observed in the peak metallicity ([Fe/H] $\approx -0.7$) between the two $L_Z$ bins, consistent with the overall uniformity of the high-$\alpha$ sequence (Fig.~\ref{fig:mdf}). Likewise, the low-$\alpha$ sample was modelled with two Gaussian components, following the procedure used for the main sample (Fig.~\ref{fig:mdf_lowafe}). The secondary component, corresponding to the LAHN population, consistently peaked at [Fe/H] $\approx -0.8$. The relative fraction compared to the accreted component was estimated to be $0.23\pm0.13$ and $0.52\pm0.24$ in $0 < L_Z < 600\ \kks$ and $-600 < L_Z < 0\ \kks$, respectively. Consequently, the large uncertainties prevent any firm conclusion regarding variation across the two $L_Z$ bins.

\begin{figure*}[h]
\centering
\vspace{0.5cm}
\begin{subfigure}{8.4cm}
    \centering
    \includegraphics[width=\textwidth]{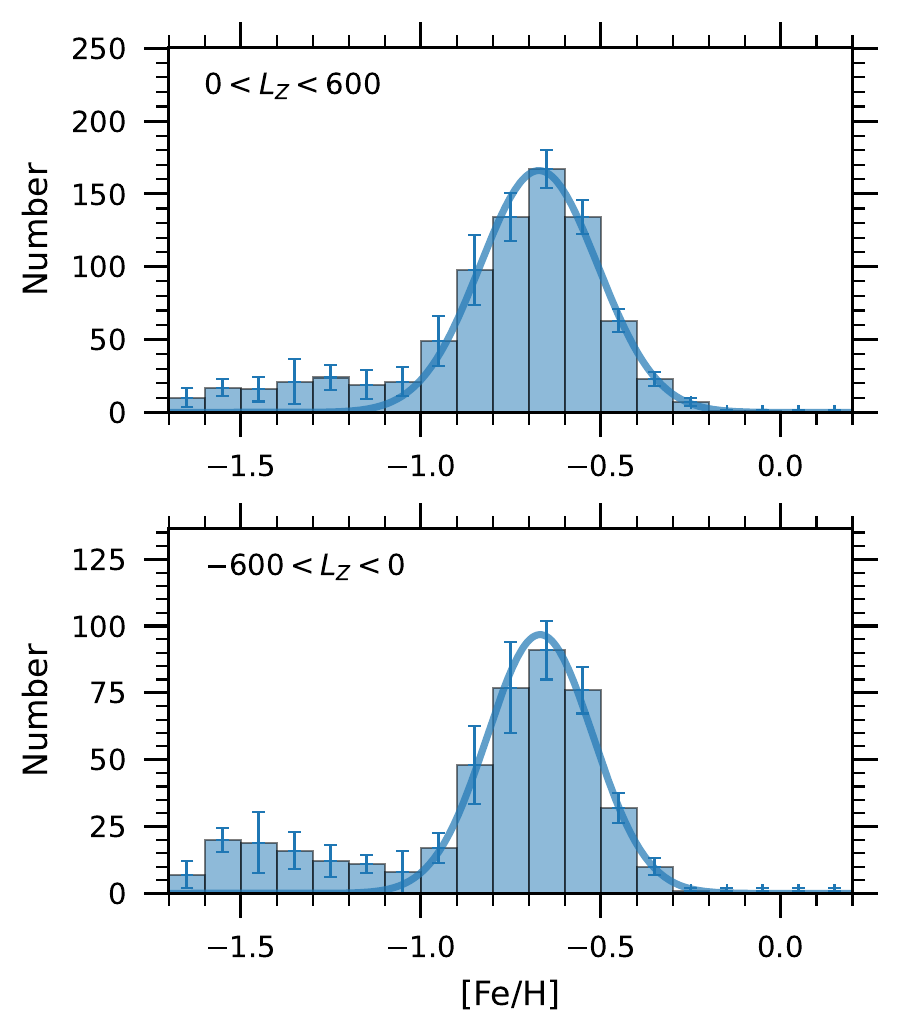}
    \caption{\textbf{High-$\alpha$ Stars}}
\end{subfigure}
\vspace{0.2cm}
\begin{subfigure}{8.4cm}
    \centering
    \includegraphics[width=\textwidth]{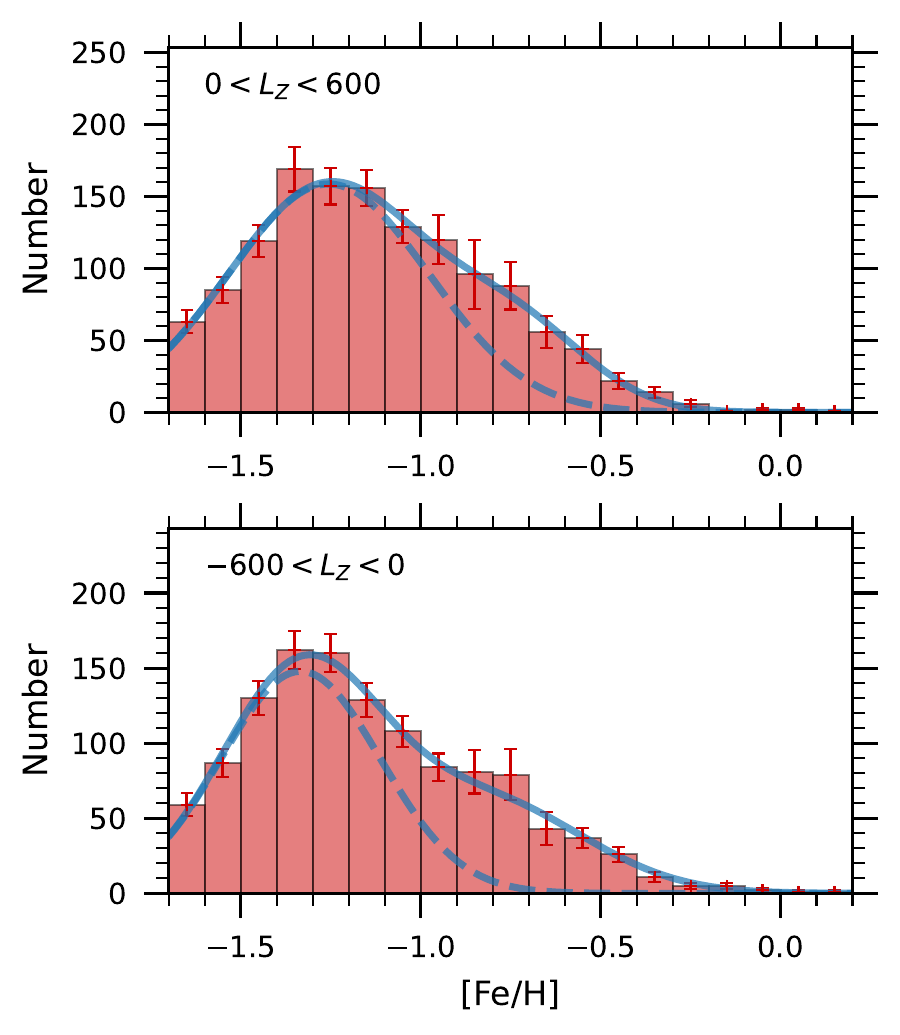}
    \caption{\textbf{Low-$\alpha$ Stars}}
\end{subfigure}
\caption{Decomposition of the metallicity distribution of the LAMOST sample, subdivided into $0 < L_Z < 600~\kks$ (top) and $-600 < L_Z < 0~\kks$ (bottom), respectively. The left panels show the case of high-$\alpha$ stars, and the right panels for low-$\alpha$ stars. The solid lines represent the best-fitting results from the same decomposition procedure as in the main samples (see Figs.~\ref{fig:mdf_highafe} and \ref{fig:mdf_lowafe}).}
\label{fig:mdf_lamost}
\end{figure*}

\end{appendix}


\begin{thebibliography}{}

\bibitem[Abdurro'uf et al.(2022)]{sdss:dr17} Abdurro'uf, Accetta, K., Aerts, C., et al.\ 2022, \apjs, 259, 35
\bibitem[Abolfathi et al.(2018)]{sdss:dr14} Abolfathi, B., Aguado, D.~S., Aguilar, G., et al.\ 2018, \apjs, 235, 42
\bibitem[Allende Prieto et al.(2008)]{allende:08} Allende Prieto, C., Sivarani, T., Beers, T.~C., et al. \ 2008, \aj, 136, 2070
\bibitem[An \& Beers(2020)]{paper1} An, D. \& Beers, T.~C.\ 2020, \apj, 897, 39
\bibitem[An \& Beers(2021a)]{paper2} An, D. \& Beers, T.~C.\ 2021a, \apj, 907, 101
\bibitem[An \& Beers(2021b)]{paper3} An, D. \& Beers, T.~C.\ 2021b, \apj, 918, 74
\bibitem[An et al.(2024)]{paper5} An, D., Beers, T.~C., \& Chiti, A.\ 2024, \apjs, 272, 20
\bibitem[An et al.(2013)]{an:13} An, D., Beers, T.~C., Johnson, J.~A., et al.\ 2013, \apj, 763, 65
\bibitem[An et al.(2023)]{paper4} An, D., Beers, T.~C., Lee, Y.~S., et al.\ 2023, \apj, 952, 66
\bibitem[An et al.(2017)]{an:17} An, D., Lee, Y.~S., Jung, J.~I., et al.\ 2017, \aj, 154, 150
\bibitem[An et al.(2011)]{an:11} An, D., Ram{\'\i}rez, S.~V., Sellgren, K., et al.\ 2011, \apj, 736, 133
\bibitem[Barnes(2002)]{barnes:02} Barnes, J.~E.\ 2002, \mnras, 333, 481
\bibitem[Barnes \& Hernquist(1996)]{barnes:96} Barnes, J.~E. \& Hernquist, L.\ 1996, \apj, 471, 115
\bibitem[Bekki \& Freeman(2003)]{bekki:03} Bekki, K. \& Freeman, K.~C.\ 2003, \mnras, 346, L11
\bibitem[Belokurov et al.(2018)]{belokurov:18} Belokurov, V., Erkal, D., Evans, N.~W., et al.\ 2018, \mnras, 478, 611
\bibitem[Belokurov et al.(2020)]{belokurov:20} Belokurov, V., Sanders, J.~L., Fattahi, A., et al.\ 2020, \mnras, 494, 3880
\bibitem[Bennett \& Bovy(2019)]{bennett:19} Bennett, M. \& Bovy, J.\ 2019, \mnras, 482, 1417
\bibitem[Bird et al.(2013)]{bird:13} Bird, J.~C., Kazantzidis, S., Weinberg, D.~H., et al.\ 2013, \apj, 773, 43
\bibitem[Blanton(2017)]{blanton:17} Blanton, M. R., Bershady, M. A., Abolfathi, B., et al. \ 2017, \aj, 154, 28
\bibitem[Bonaca et al.(2017)]{bonaca:17} Bonaca, A., Conroy, C., Wetzel, A., et al.\ 2017, \apj, 845, 101
\bibitem[Bovy(2015)]{bovy:15} Bovy, J.\ 2015, \apjs, 216, 29
\bibitem[Brook et al.(2004)]{brook:04} Brook, C.~B., Kawata, D., Gibson, B.~K., et al.\ 2004, \apj, 612, 894
\bibitem[Brook et al.(2012)]{brook:12} Brook, C.~B., Stinson, G., Gibson, B.~K., et al.\ 2012, \mnras, 419, 771
\bibitem[Buck(2020)]{buck:20} Buck, T.\ 2020, \mnras, 491, 5435
\bibitem[Buder et al.(2025)]{buder:25} Buder, S., Kos, J., Wang, X.~E., et al.\ 2025, \pasa, 42, e051
\bibitem[Chabrier(2003)]{chabrier:03} Chabrier, G.\ 2003, \pasp, 115, 763
\bibitem[Ciuc{\u{a}} et al.(2024)]{ciuca:24} Ciuc{\u{a}}, I., Kawata, D., Ting, Y.-S., et al.\ 2024, \mnras, 528, L122
\bibitem[Claeyssens et al.(2023)]{claeyssens:23} Claeyssens, A., Adamo, A., Richard, J., et al.\ 2023, \mnras, 520, 2180
\bibitem[Conroy et al.(2022)]{conroy:22} Conroy, C., Weinberg, D.~H., Naidu, R.~P., et al.\ 2022, submitted to OJAp (arXiv:2204.02989)
\bibitem[Cui et al.(2012)]{cui:12} Cui, X.~Q., Zhao, Y.~H., Chu, Y.~Q., et al. \ 2012, RAA, 12, 1197
\bibitem[Dawson et al.(2013)]{dawson:13} Dawson, K.~S., Schlegal, D.~J., Ahn, C., et al. \ 2013, \aj, 145, 10
\bibitem[Deason et al.(2018)]{deason:18} Deason, A.~J., Belokurov, V., Koposov, S.~E., et al.\ 2018, \apjl, 862, L1
\bibitem[Di Matteo et al.(2019)]{dimatteo:19} Di Matteo, P., Haywood, M., Lehnert, M.~D., et al.\ 2019, \aap, 632, A4
\bibitem[Escala et al.(2018)]{escala:18} Escala, I., Wetzel, A., Kirby, E.~N., et al.\ 2018, \mnras, 474, 2194
\bibitem[Ferland et al.(2013)]{ferland:13} Ferland, G.~J., Porter, R.~L., van Hoof, P.~A.~M., et al.\ 2013, \rmxaa, 49, 137
\bibitem[F{\"o}rster Schreiber et al.(2009)]{forsterschreiber:09} F{\"o}rster Schreiber, N.~M., Genzel, R., Bouch{\'e}, N., et al.\ 2009, \apj, 706, 1364
\bibitem[Fujii et al.(2021)]{fujii:21} Fujii, M.~S., Saitoh, T.~R., Hirai, Y., et al.\ 2021, \pasj, 73, 1074
\bibitem[Gaia Collaboration et al.(2023)]{gaia:dr3} Gaia Collaboration, Vallenari, A., Brown, A.~G.~A., et al.\ 2023, \aap, 674, A1
\bibitem[Grand et al.(2020)]{grand:20} Grand, R.~J.~J., Kawata, D., Belokurov, V., et al.\ 2020, \mnras, 497, 1603
\bibitem[Helmi et al.(2018)]{helmi:18} Helmi, A., Babusiaux, C., Koppelman, H.~H., et al.\ 2018, \nat, 563, 85
\bibitem[Hirai et al.(2022)]{hirai:22} Hirai, Y., Beers, T.~C., Chiba, M., et al.\ 2022, \mnras, 517, 4856
\bibitem[Hirai et al.(2021)]{hirai:21} Hirai, Y., Fujii, M.~S., \& Saitoh, T.~R.\ 2021, \pasj, 73, 1036
\bibitem[Hirai et al.(2024)]{hirai:24} Hirai, Y., Kirby, E.~N., Chiba, M., et al.\ 2024, \apj, 970, 105
\bibitem[Hirai \& Saitoh(2017)]{hirai:17} Hirai, Y. \& Saitoh, T.~R.\ 2017, \apjl, 838, L23
\bibitem[Ivezi{\'c} et al.(2019)]{ivezic:19} Ivezi{\'c}, {\v{Z}}., Kahn, S.~M., Tyson, J.~A., et al.\ 2019, \apj, 873, 111
\bibitem[Johnson(2019)]{johnson:19} Johnson, J.~A.\ 2019, Science, 363, 474
\bibitem[Johnson \& Pilachowski(2010)]{johnson:10} Johnson, C.~I. \& Pilachowski, C.~A.\ 2010, \apj, 722, 1373
\bibitem[Kennicutt(1998)]{kennicutt:98} Kennicutt, R.~C.\ 1998, \araa, 36, 189
\bibitem[Khoperskov et al.(2023)]{khoperskov:23} Khoperskov, S., Minchev, I., Libeskind, N., et al.\ 2023, \aap, 677, A89
\bibitem[Lee et al.(2023)]{lee:23} Lee, A., Lee, Y.~S., Kim, Y.~K., et al.\ 2023, \apj, 945, 56
\bibitem[Lee et al.(2011)]{lee:11} Lee, Y. S., Beers, T. C., Allende Prieto, C., et al. \ 2011, \aj, 141, 90
\bibitem[Lee et al.(2015)]{lee:15} Lee, Y.~S., Beers, T.~C., Carlin, J.~L., et al. \ 2015, \aj, 150, 187
\bibitem[Lee et al.(2013)]{lee:13} Lee, Y.~S., Beers, T.~C., Masseron, T., et al. \ 2013, \aj, 146, 132
\bibitem[Lee et al.(2008a)]{lee:08a} Lee, Y.~S., Beers, T.~C., Sivarani, T., et al. \ 2008a, \aj, 136, 2022
\bibitem[Lee et al.(2008b)]{lee:08b} Lee, Y.~S., Beers, T.~C., Sivarani, T., et al. \ 2008b, \aj, 136, 2050
\bibitem[Lee et al.(1999)]{lee:99} Lee, Y.-W., Joo, J.-M., Sohn, Y.-J., et al.\ 1999, \nat, 402, 55
\bibitem[Luo et al.(2015)]{luo:15} Luo, A.-L., Zhao, Y.-H., Zhao, G., et al. \ 2015, RAA, 15, 1095
\bibitem[Lian \& Luo(2024)]{lian:24} Lian, J. \& Luo, L.\ 2024, \apjl, 960, L10
\bibitem[Lindegren et al.(2021)]{lindegren:21} Lindegren, L., Bastian, U., Biermann, M., et al.\ 2021, \aap, 649, A4
\bibitem[Liu et al.(2024)]{liu:24} Liu, Z., Silverman, J.~D., Daddi, E., et al.\ 2024, \apj, 968, 15
\bibitem[Ma et al.(2017)]{ma:17} Ma, X., Hopkins, P.~F., Wetzel, A.~R., et al.\ 2017, \mnras, 467, 2430
\bibitem[Majewski et al.(2017)]{majewski:17} Majewski, S.~R., Schiavon, R.~P., Frinchaboy, P.~M., et al.\ 2017, \aj, 154, 94
\bibitem[Matsuno et al.(2024)]{matsuno:24} Matsuno, T., Amarsi, A.~M., Carlos, M., et al.\ 2024, \aap, 688, A72
\bibitem[McCluskey et al.(2024)]{mccluskey:24} McCluskey, F., Wetzel, A., Loebman, S.~R., et al.\ 2024, \mnras, 527, 6926
\bibitem[McMillan(2017)]{mcmillan:17} McMillan, P.~J.\ 2017, \mnras, 465, 76
\bibitem[Mihos \& Hernquist(1996)]{mihos:96} Mihos, J.~C. \& Hernquist, L.\ 1996, \apj, 464, 641
\bibitem[Moreno et al.(2015)]{moreno:15} Moreno, J., Torrey, P., Ellison, S.~L., et al.\ 2015, \mnras, 448, 1107
\bibitem[Mowla et al.(2022)]{mowla:22} Mowla, L., Iyer, K.~G., Desprez, G., et al.\ 2022, \apjl, 937, L35
\bibitem[Myeong et al.(2022)]{myeong:22} Myeong, G.~C., Belokurov, V., Aguado, D.~S., et al.\ 2022, \apj, 938, 21
\bibitem[Naidu et al.(2021)]{naidu:21} Naidu, R.~P., Conroy, C., Bonaca, A., et al.\ 2021, \apj, 923, 92
\bibitem[Onken et al.(2024)]{smss:dr4} Onken, C.~A., Wolf, C., Bessell, M.~S., et al.\ 2024, \pasa, 41, e061
\bibitem[Pancino et al.(2000)]{pancino:00} Pancino, E., Ferraro, F.~R., Bellazzini, M., et al.\ 2000, \apjl, 534, L83
\bibitem[Papovich et al.(2015)]{papovich:15} Papovich, C., Labb{\'e}, I., Quadri, R., et al.\ 2015, \apj, 803, 26
\bibitem[Pierce-Price et al.(2000)]{pierceprice:00} Pierce-Price, D., Richer, J.~S., Greaves, J.~S., et al.\ 2000, \apjl, 545, L121
\bibitem[Reid et al.(2014)]{reid:14} Reid, M.~J., Menten, K.~M., Brunthaler, A., et al.\ 2014, \apj, 783, 130
\bibitem[Renaud et al.(2021)]{renaud:21} Renaud, F., Agertz, O., Read, J.~I., et al.\ 2021, \mnras, 503, 5846
\bibitem[Renaud et al.(2014)]{renaud:14} Renaud, F., Bournaud, F., Kraljic, K., et al.\ 2014, \mnras, 442, L33
\bibitem[Renaud et al.(2022)]{renaud:22} Renaud, F., Segovia Otero, {\'A}., \& Agertz, O.\ 2022, \mnras, 516, 4922
\bibitem[Rockosi et al.(2022)]{rockosi:22} Rockosi, C. M., Sun Lee, Y., Morrison, H. L., et al. \ 2022, \apjs, 259, 60
\bibitem[Saitoh(2017)]{saitoh:17} Saitoh, T.~R.\ 2017, \aj, 153, 85
\bibitem[Saitoh et al.(2008)]{saitoh:08} Saitoh, T.~R., Daisaka, H., Kokubo, E., et al.\ 2008, \pasj, 60, 667
\bibitem[Saitoh et al.(2009)]{saitoh:09} Saitoh, T.~R., Daisaka, H., Kokubo, E., et al.\ 2009, \pasj, 61, 481
\bibitem[Sales et al.(2009)]{sales:09} Sales, L.~V., Helmi, A., Abadi, M.~G., et al.\ 2009, \mnras, 400, L61
\bibitem[Schmidt(1959)]{schmidt:59} Schmidt, M.\ 1959, \apj, 129, 243
\bibitem[Sch{\"o}nrich(2012)]{schonrich:12} Sch{\"o}nrich, R.\ 2012, \mnras, 427, 274
\bibitem[Sch{\"o}nrich et al.(2010)]{schonrich:10} Sch{\"o}nrich, R., Binney, J., \& Dehnen, W.\ 2010, \mnras, 403, 1829
\bibitem[Smolinski et al.(2011)]{smolinski:11} Smolinski, J. P., Lee, Y. S., Beers, T. C., et al. \ 2011, \aj, 141, 89
\bibitem[Sparre et al.(2017)]{sparre:17} Sparre, M., Hayward, C.~C., Feldmann, R., et al.\ 2017, \mnras, 466, 88
\bibitem[Stott et al.(2016)]{stott:16} Stott, J.~P., Swinbank, A.~M., Johnson, H.~L., et al.\ 2016, \mnras, 457, 1888
\bibitem[Tsukui et al.(2025)]{tsukui:25} Tsukui, T., Wisnioski, E., Bland-Hawthorn, J., et al.\ 2025, \mnras, 540, 4, 3493
\bibitem[Xiang et al.(2025)]{xiang:25} Xiang, M., Rix, H.-W., Yang, H., et al.\ 2025, Nature Astronomy, 9, 101
\bibitem[Yanny et al.(2009)]{yanny:09} Yanny, B., Newberg, H.~J., Johnson, J.~A., et al. \ 2009, \aj, 137, 4377
\bibitem[York et al.(2000)]{york:00} York, D.~G., Adelman, J., Anderson, J.~E., et al. \ 2000, \aj, 120, 1579
\bibitem[Yu et al.(2021)]{yu:21} Yu, S., Bullock, J.~S., Klein, C., et al.\ 2021, \mnras, 505, 889

\end{thebibliography}
\end{document}